\documentclass[10pt,journal,compsoc]{IEEEtran}
\ifCLASSOPTIONcompsoc
  \usepackage[nocompress]{cite}
\else
  \usepackage{cite}
\fi

\usepackage{times}
\usepackage{xcolor}
\usepackage{soul}
\usepackage[utf8]{inputenc}
\usepackage[small,tableposition=bottom]{caption}
\usepackage[labelformat=simple]{subcaption}
\usepackage{booktabs} % For formal tables
\usepackage{amsmath}
\usepackage{graphicx}
\usepackage{float}
\usepackage{enumitem}
\usepackage{multirow}
\usepackage{bm}
\usepackage{mathrsfs}
\usepackage{amssymb}
\usepackage[linesnumbered,boxed,ruled,commentsnumbered]{algorithm2e}
\usepackage[normalem]{ulem} % use normalem to protect \emph
\newcommand\hll{\bgroup\markoverwith
  {\textcolor{yellow}{\rule[-.5ex]{2pt}{2.5ex}}}}

\graphicspath{ {./figures/} }
\ifCLASSINFOpdf
\else
\fi
\begin{document}
\title{
% Aspect-Level Deep Collaborative Filtering via Heterogeneous Information Networks
Deep Collaborative Filtering with Multi-Aspect Information in Heterogeneous Networks
}

\author{Chuan Shi,~\IEEEmembership{Member,~IEEE,}
        Xiaotian Han,
        Li Song, 
        Xiao Wang*,
        Senzhang Wang,
        Junping Du,
        \\Philip S. Yu,~\IEEEmembership{Fellow,~IEEE,}
\IEEEcompsocitemizethanks{\IEEEcompsocthanksitem Chuan Shi, Xiaotian Han, Li Song, Xiao Wang and Junping Du are with the School of Computer Science, Beijing University of Posts and Telecommunications, Beijing,
China. E-mail: shichuan@bupt.edu.cn; hanxiaotian.h@gmail.com; song200626@gmail.com; xiaowang@bupt.edu.cn; junpingdu@126.com;
% \protect\\
\IEEEcompsocthanksitem Senzhang Wang is with Nanjing University of Aeronautics and Astronautics. E-mail: szwang@nuaa.edu.cn
\IEEEcompsocthanksitem Philip S. Yu is with University of Illinois at Chicago, Institude for Data Science, Tsinghua University. E-mail: psyu@cs.uic.edu

\IEEEcompsocthanksitem * The corresponding author is Xiao Wang.
}% <-this % stops a space
% \thanks{Manuscript received April 19, 2005; revised August 26, 2015.}
}

\IEEEtitleabstractindextext{%
\begin{abstract}
Recently, recommender systems play a pivotal role in alleviating the problem of information overload. 
Latent factor models have been widely used for recommendation. Most existing latent factor models mainly utilize the interaction information between users and items, although some recently extended models utilize some auxiliary information to learn a unified latent factor for users and items.  The unified latent factor only represents the characteristics of users and the properties of items from the aspect of purchase history. However, the characteristics of users and the properties of items may stem from different aspects, e.g., the brand-aspect and category-aspect of items. 
Moreover, the latent factor models usually use the shallow projection, which cannot capture the characteristics of users and items well. Deep neural network has shown tremendous potential to model the non-linearity relationship between users and items.  
It can be used to replace shallow projection to model the complex correlation between users and items. 
% This has been verified in speech recognition, computer vision and natural language processing.
In this paper, we propose a Neural network based Aspect-level Collaborative Filtering model (NeuACF) to exploit different aspect latent factors. Through modelling the rich object properties and relations in recommender system as a heterogeneous information network, NeuACF first extracts different aspect-level similarity matrices of users and items respectively through different meta-paths, and then feeds an elaborately designed deep neural network with these matrices to learn aspect-level latent factors. Finally, the aspect-level latent factors are fused for the top-N recommendation. 
Moreover, to fuse information from different aspects more effectively, we further propose NeuACF++ to fuse aspect-level latent factors with self-attention mechanism. 
Extensive experiments on three real world datasets show that NeuACF and NeuACF++ significantly outperform both existing latent factor models and recent neural network models.
\end{abstract}

}

\maketitle

% is not selected <OR> if conference mode is selected - because compsoc
% conference papers position the abstract like regular (non-compsoc)
\IEEEdisplaynontitleabstractindextext

% \fi
\IEEEpeerreviewmaketitle

% the following package is optional:
%\usepackage{latexsym} 

% Following comment is from ijcai97-submit.tex:
% The preparation of these files was supported by Schlumberger Palo Alto
% Research, AT\&T Bell Laboratories, and Morgan Kaufmann Publishers.
% Shirley Jowell, of Morgan Kaufmann Publishers, and Peter F.
% Patel-Schneider, of AT\&T Bell Laboratories collaborated on their
% preparation.

% These instructions can be modified and used in other conferences as long
% as credit to the authors and supporting agencies is retained, this notice
% is not changed, and further modification or reuse is not restricted.
% Neither Shirley Jowell nor Peter F. Patel-Schneider can be listed as
% contacts for providing assistance without their prior permission.

% To use for other conferences, change references to files and the
% conference appropriate and use other authors, contacts, publishers, and
% organizations.
% Also change the deadline and address for returning papers and the length and
% page charge instructions.
% Put where the files are available in the appropriate places.

% Single author syntax
% \author{Jêröme Lang\\ 
% Laboratoire d'Analyse et Modélisation des Systèmes pour l'Aide à la Décision (LAMSADE)  \\
% pcchair@ijcai-18.org}

% Multiple author syntax (remove the single-author syntax above and the \iffalse ... \fi here)

\section{Introduction}
\label{intro}
Currently the overloaded online information overwhelms users. In order to tackle the problem, Recommender Systems (RS) are widely employed to guide users in a personalized way of discovering products or services they might be interested from a large number of possible alternatives. 
Recommender systems are essential for e-commerce companies to provide users a personalized recommendation of products, and thus most e-commerce companies like Amazon and Alibaba are in an urgent need to build more effective recommender systems to improve user experience. 
Due to its importance in practice, recommender systems have been attracting remarkable attention to both industry and academic research community. 

Collaborative Filtering (CF)~\cite{hu2008collaborative} is one of the most popular methods for recommendation, whose basic assumption is that people who share similar purchase in the past tend to have similar choices in the future. In order to exploit users' similar purchase preference, latent factor models (e.g., matrix factorization)~\cite{koren2009matrix,koren2008factorization} have been proposed, which usually factorize the user-item interaction matrix (e.g., rating matrix) into two low-rank user-specific and item-specific factors, and then use the low-rank factors to make predictions. Since latent factor models may suffer from data sparsity, many extended latent factor models integrate auxiliary information into the matrix factorization framework, such as social recommendation~\cite{ma2008sorec} and heterogeneous network based recommendation~\cite{shi2016integrating}. Recently, with the surge of deep learning, deep neural networks are also employed to deeply capture the latent features of users and items for recommendation. NeuMF~\cite{he2017neural} replaces the inner product operations in matrix factorization with a multi-layer feed-forward neural network to capture the non-linear relationship between users and items. DMF~\cite{xue2017deep} uses the rating matrix directly as the input and maps user and items into a common low-dimensional space via a deep neural network.

\begin{figure}
  \centering
  \begin{subfigure}[t]{0.16\textwidth}
      \centering
      \includegraphics[height=1.in, width=1.0in]{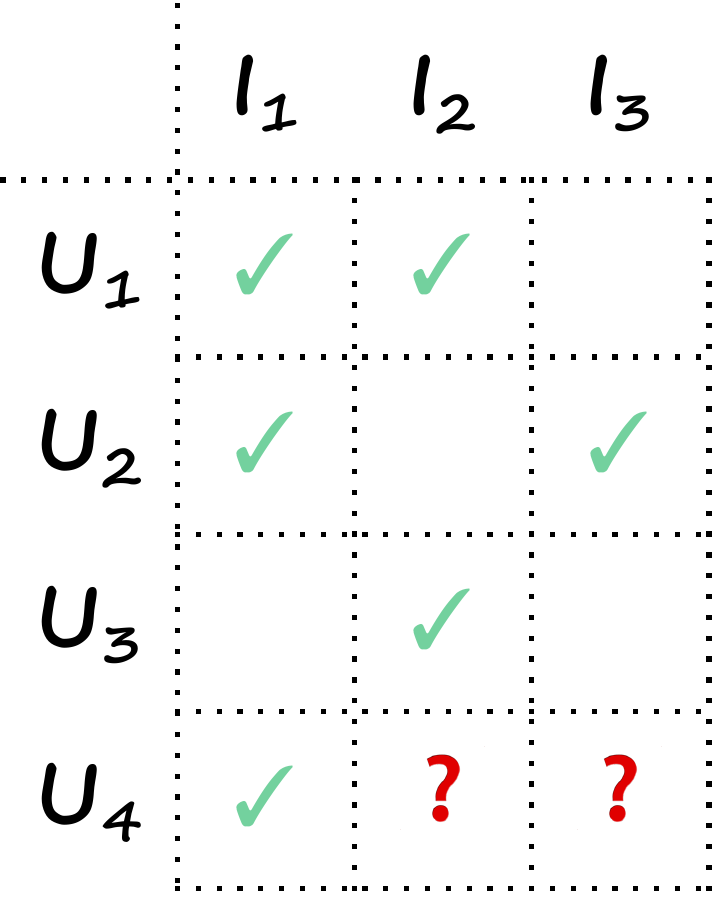}
      \caption{User-Item purchase relation}
      \label{fig:um}
  \end{subfigure}
  \centering
  \begin{subfigure}[t]{0.15\textwidth}
      \centering
      \includegraphics[height=1.in, width=1.0in]{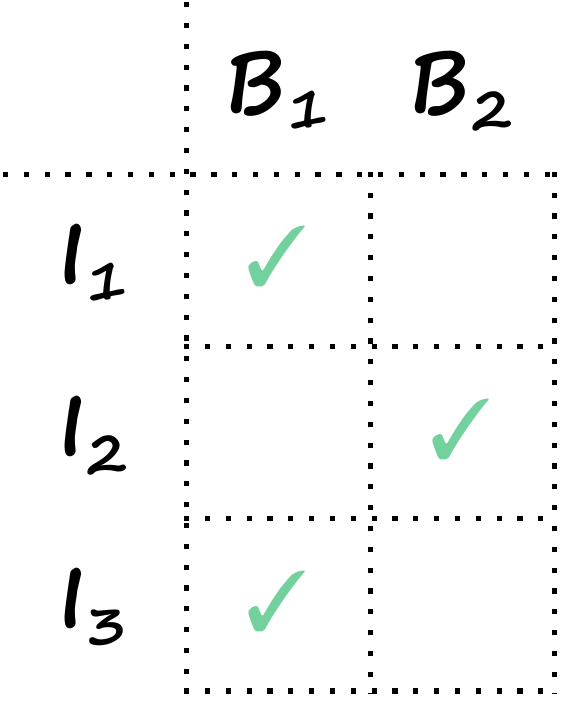}
      \caption{Item-Brand relation}
      \label{fig:md}
  \end{subfigure}
  \centering
  \begin{subfigure}[t]{0.16\textwidth}
      \centering
      \includegraphics[height=1.0in, width=0.9in]{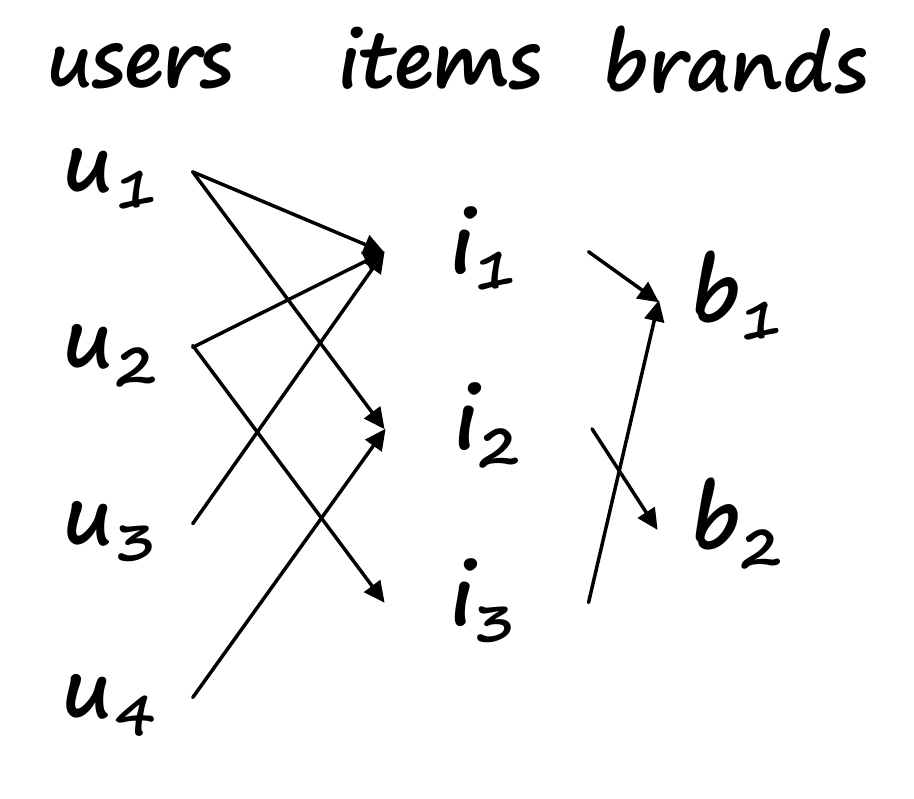}
      \caption{heterogeneous network}
      \label{fig:umd}
  \end{subfigure}
  \caption{A toy example of aspect-level interactions between users and items.}
  \label{fig:toy}
\end{figure}

Although these latent factor models achieve good performance, they usually only capture the information of users' purchase history. Existing models usually focus on extracting latent factors of users and items through their interaction information from ratings, which only reflects user preferences and item characteristics from one aspect, i.e., purchase history. However, the latent factors of users and items usually stem from different aspects in real applications. Particularly, in social media with rich information, user preferences and item characteristics may reflect in many aspects besides rating interactions, e.g., item features, and other interactions between users.  
These aspect-level features can more comprehensively reflect user preferences and item characteristics. Thus it is very valuable for the latent factor models to exploit latent features of users and items from different aspects. 
Figure~\ref{fig:toy} shows a toy example of our idea. A green check mark indicates that the user purchased the corresponding item in the past. A question mark means that the interaction information is unknown. If we only exploit the interaction matrix (illustrating purchase history) in Figure~\ref{fig:um}, we may infer that user $U_4$ will purchase item $I_2$ and $I_3$. However, when considering the item brand information shown in Figure~\ref{fig:md}, we may find item $I_3$ is a better recommendation to $U_4$ because items $I_1$ and $I_3$ belong to the same brand $B_1$.

Although it is promising to comprehensively utilize multiple aspect-level latent features of users and items, it still faces the following two challenges. (1) How to extract different aspect-level features: A systematic method is needed to effectively organize the different types of objects and interactions in recommender systems, and extract different aspect-level features. The extracted aspect-level features should reflect different aspects of users preferences and embody rich semantics. 
(2) How to learn latent factors from different aspects. Even if we can extract different aspect-level features, it is still not easy to learn their latent factors. Matrix factorization may not be a good option as it only learns the  shallow  factors. 
Deep neural network (DNN), which is able to learn the highly nonlinear representations of users and items, is a promising method. However, the current DNN structure lacks of feature fusing mechanism, which cannot be directly applied to our problem.
(3) How to fuse latent factors from different aspects effectively. Since the different aspect-level factors only represent aspect-level characteristics of user/item, we need to fuse them effectively. Although deep neural network is a promising method, we still need to design a proper neural network structure and a feature fusing mechanism for our problem settings.

In this paper, to address the challenges above, we propose a novel Neural network based Aspect-level Collaborative Filtering model (NeuACF). NeuACF can effectively model and fuse different aspect-level latent factors which represent the user preferences and item characteristics from different aspects. Particularly, the objects and interactions of different types in recommender systems are first organized as a Heterogeneous Information Network (HIN) ~\cite{shi2017survey}. Meta-paths~\cite{sun2011pathsim}, relation sequences connecting objects, are then employed to extract aspect-level features of users and items. As an example shown in Figure~\ref{fig:umd}, we can extract the latent factors of users from the aspect of purchase history with the $User$-$Item$-$User$ path, which is usually analyzed by existing latent factor models. Meanwhile, we can also extract the latent factors from the aspect of brand preference with the $User$-$Item$-$Brand$-$Item$-$User$ path. Furthermore, we design a delicate deep neural network to learn different aspect-level latent factors for users and items and utilize an attention mechanism to effectively fuse them for the top-N recommendation. Note that, different from those hybrid recommendation models~\cite{wang2015collaborative} that focus on the rating information with the auxiliary information, NeuACF treats different aspect-level latent factors extracted from meta-paths equally, and automatically determines the importance of these aspects. NeuACF is also different from those HIN based methods~\cite{zhao2017meta} in its deep model and fusing mechanism. 
Concretely, a delicately designed attention network is used to fuse aspect-level latent factors. Comparing to the above attention method, we further propose NeuACF++ to fuse aspect information with self-attention mechanism which considers different aspect-level latent factors and learns the attention values simultaneously.
Extensive experiments illustrate the effectiveness of NeuACF and NeuACF++, as well as the traits of aspect-level latent factors.

Our main contributions of this paper are summarized as follows.
\begin{itemize}
  \item To leverage the different aspect-level information of HIN, we design a meta-path based method to capture the aspect-level latent factors of users and items from the similarity matrix obtained from the HIN.
  
  \item We propose the NeuACF with deep neural network to learn different aspect-level latent factors and integrate these latent factors with attention mechanism for top-N recommendation, since aspect-level information reflects the characteristics of users and the properties of items more precisely. Moreover, the self-attention mechanism is employed to fuse aspect-level latent factors in our proposed method NeuACF++.
  
  \item We preform extensive experiments and provide tremendous analysis to illustrate the effectiveness of NeuACF and NeuACF++. 
\end{itemize}

  The rest of this paper is organized as follows. Section \ref{sec:related} reviews the related work. Section \ref{sec:preli} summarizes the related work. Section \ref{sec:model} introduces the NeuACF model and NeuACF++ model in details. Section \ref{sec:experiment} presents and analyzes the experimental results. And Section \ref{sec:conclusion} concludes this paper.

\section{Related Work}
\label{sec:related}
In this section, we provide a background to our work, and review the relevant works.

\subsection{ Collaborative Filtering}
\label{sec:implicitCF}
Traditional recommendation works mainly adopt collaborative filtering (CF) methods to utilize historical interactions for recommendation~\cite{schafer2007collaborative,he2016fast,koren2008factorization,liu2017learning}. As the most popular approach among various CF techniques, matrix factorization (MF) has shown its effectiveness and efficiency in many applications~\cite{koren2009matrix,shi2012adaptive}. MF factorizes the user-item interaction matrix into two low-dimension user-specific and item-specific matrices, and then utilizes the factorized matrices for predictions~\cite{koren2015advances}. In recent years, many variants of MF, such as SVD~\cite{koren2008factorization}, weighted regularized matrix factorization~\cite{hu2008collaborative}, and probabilistic matrix factorization~\cite{mnih2008probabilistic} have been proposed. 
SVD reconstructs the rating matrix only through the observed user-item interactions. 
Weighted regularized matrix factorization (WR-MF) extends MF by using regularization to prevent over-fitting and to increase the impact of positive feedback.
Probabilistic matrix factorization (PMF) models the user preference matrix as a product of two lower-rank user and item matrices. The user and item feature vectors are computed by a probabilistic linear model with Gaussian observation distribution.
Bayesian personalized ranking (BPR)~\cite{rendle2009bpr} is a generic optimization criterion and learning algorithm for implicit CF and has been widespreadly adopted in many related domains~\cite{dave2018neural,he2016vbpr,niu2018neural,zhang2016trust}. 

\subsection{ Neural Networks for Recommendation }
\label{sec:nnRec}
Recently, neural network has shown its potential in non-linear transformations and been successfully applied in many data mining tasks~\cite{hoff2002latent,yan2007graph}. The neural network has been proven to be capable of approximating any continuous function~\cite{hornik1989multilayer}. 
The pioneer work proposes a two-layers Restricted Boltzmann Machines (RBMs) to model user-item interactions~\cite{salakhutdinov2007restricted}. 
In addition, autoencoders have been applied to learn user and item vectors for recommendation systems~\cite{li2015deep,sedhain2015autorec,strub2015collaborative}.
To overcome the limitation of autoencoders and increase the generalization ability, denoising autoencoders (DAE) have been applied to learn user and item vectors from intentionally corrupted inputs~\cite{li2015deep,strub2015collaborative}.
Cheng et al.~\cite{cheng2016wide} combine the benefits of memorization and generalization for recommender systems by jointly training wide linear models and deep neural networks. 
Compared to Wide \& Deep model,  Guo et al.~\cite{guo2017deepfm} propose the DeepFM model that integrates the architectures of factorization machine (FM) and deep neural networks (DNN). This architecture models low-order feature interactions and high-order feature interactions simultaneously. 
He et al.~\cite{he2017neural} present a neural network architecture to model latent features of users and items and devise a general neural collaborative filtering (NCF) framework based on neural networks. In addition, NCF leverages a multi-layer perceptron to learn the user-item interaction function instead of the traditional inner product. 
He et al.~\cite{he2017neuralFM} propose the neural factorization machine (NFM) model for recommendation. This model combines the linearity of FM in modeling second-order feature interactions and the non-linearity of neural network to model higher-order feature interactions. 
Xue et al.~\cite{xue2017deep} propose a deep matrix factorization model (DMF) with a neural network that maps the users and items into a common low-dimensional space with non-linear projections. The training matrix includes both explicit ratings and non-preference implicit feedback. 
The recently proposed convolutional NCF ~\cite{he2018outer} utilizes outer product above the embedding layer results and 2D convolution layers for learning joint representation of user-item pairs. 

\subsection{Exploiting Heterogeneous Information for Recommendation}
\label{sec:hinRec}
To overcome the sparsity of the ratings, additional data are integrated into recommendation systems, such as social matrix factorization with social relations~\cite{ma2008sorec} and topicMF with item contents or reviews text ~\cite{bao2014topicmf}. 
Recently, graph data\cite{wang2017learning} shows its strong potential for many data mining tasks.
There are also many works exploring the graph data for recommendation~\cite{shi2015semantic, zhang2017joint} or web search~\cite{wang2012multimodal}.
As one of the most important methods to model the graph data,
heterogeneous information network~\cite{shi2017survey} can naturally characterize the different relations between different types and objects. Then several path based similarity measures are proposed to evaluate the similarity of objects in heterogeneous information network~\cite{sun2011pathsim,lao2010relational,shi2014hetesim}. After that, many HIN based recommendation methods have been proposed to integrate auxiliary information. 
Feng et al.~\cite{feng2012incorporating} propose a method to learn the weights of different types of nodes and edges, which can alleviate the cold start problem by utilizing heterogeneous information contained in social tagging system. 
Furthermore, meta-path is applied to recommender systems to integrate different semantic information ~\cite{liu2014meta}. 
In order to take advantage of the heterogeneity of relationship in information networks, Yu et al.~\cite{yu2014personalized} propose to diffuse user preferences along different meta-paths in information networks. 
Luo et al.~\cite{luo2014hete} demonstrate that multiple types of relations in heterogeneous social network can mitigate the data sparsity and cold start problems. 
Shi et al.~\cite{shi2015semantic} design a novel SemRec method to integrate all kinds of information contained in recommender system using weighted HIN and meta-paths. 
Zhang et al.~\cite{zhang2017joint} propose a joint representation learning (JRL) framework for top-N recommendation by integrating different latent representations.

Most existing latent factor models mainly utilize the rating information between users and items, but ignore the aspect information of users and items.
In this paper, we extract different aspect similarity matrices through different meta-paths which characterize the specific aspect information. Then, we delicately design a deep neural network to learn the latent factors of users and items. After that, we utilize attention mechanism to fuse those aspect-level latent factors for top-N recommendation. 

\section{Preliminaries}
\label{sec:preli}
\subsection{ Latent Factor Model }
The latent factor model has been widely studied in recommender systems. Its basic idea is to map users and items to latent factors and use these factors for recommendation. The representative works are Matrix Factorization (MF)~\cite{koren2009matrix}, PMF~\cite{mnih2008probabilistic} and SVD++~\cite{koren2008factorization}. 
Taking MF for example, the objective function of MF in Equation~\ref{equ:mf} aims to minimize the following regularized squared loss on the observed ratings: 
\begin{equation}
  \label{equ:mf}
  \mathop{\arg\min}_{\bm{u},\bm{v}}\sum_i \sum_j(\bm{R}_{i,j} - \bm{u}_i^T\bm{v}_j)^2 + \lambda \left( \sum_i||\bm{u}_i||_2^2 +  \sum_j||\bm{v}_j||_2^2 \right),
\end{equation}
where $\bm{u}_i$ and $\bm{v}_j$ denote the latent factors of user $U_i$ and item $I_j$, $\bm{R}_{i,j}$ denote the user $U_i$ rating score to item $I_j$ and the $\lambda$ controls the strength of regularization, which is usually a $L$-2 norm aiming to prevent overfitting. 

Based on this basic MF framework, many extended latent factor models have been proposed through adding some auxiliary information, such as social recommendation~\cite{ma2008sorec} and heterogeneous network based recommendation~\cite{shi2015semantic}. The limitation of existing latent factor models is that the latent factors are mainly extracted from one aspect, i.e., the rating matrix. However, some other more fine-grained aspect-level user-item interaction information is largely ignored, although such information is also useful.

\subsection{Heterogeneous Information Network}
\begin{figure}
  \centering
  \begin{subfigure}[t]{0.2\textwidth}
      \centering
      \includegraphics[height=0.85in, width=1.0in]{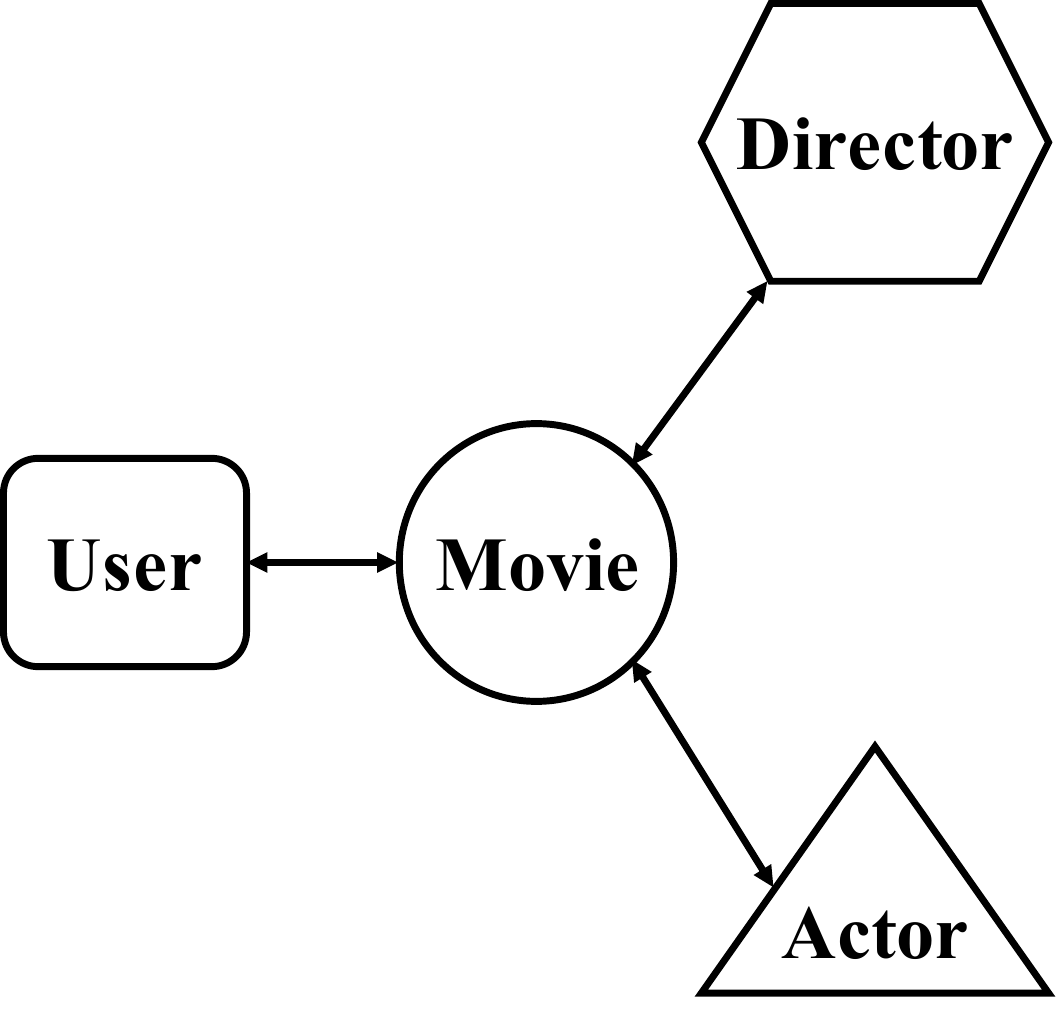}
      \caption{MovieLens}
      \label{fig:ml}
  \end{subfigure}%
~
  \centering
  \begin{subfigure}[t]{0.2\textwidth}
      \centering
      \includegraphics[height=0.9in, width=1.0in]{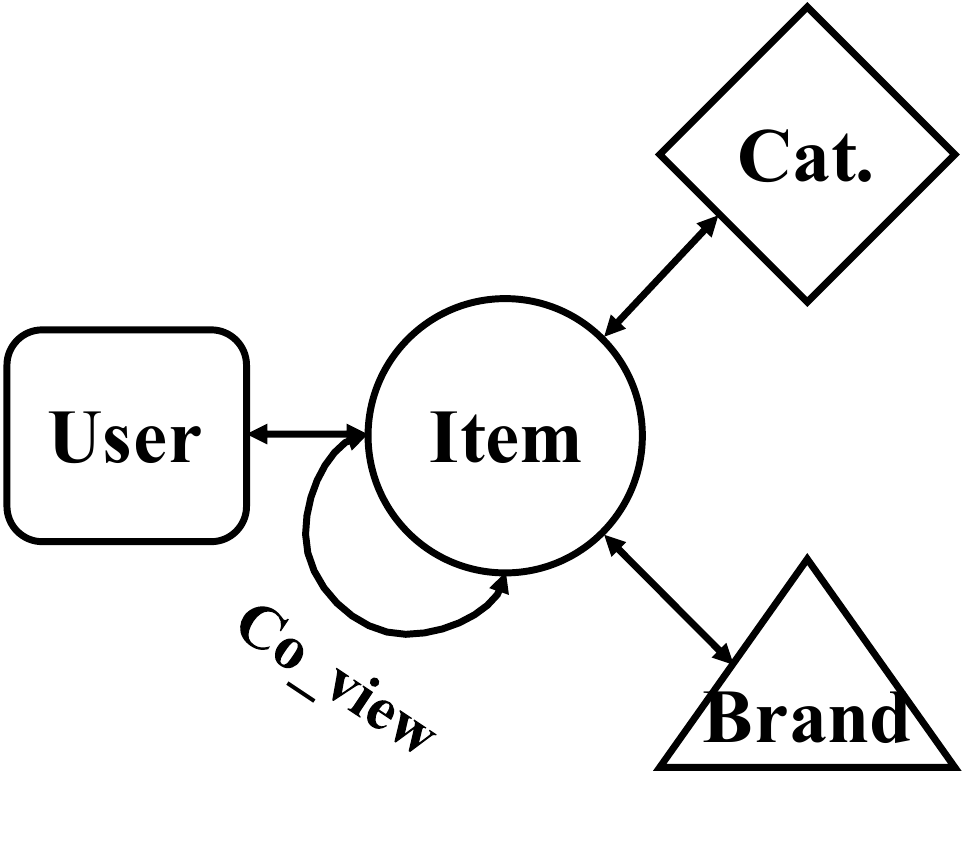}
      \caption{Amazon}
      \label{fig:amazon}
  \end{subfigure}%
  \caption{Network schema of HINs for the experimental datasets.}
  \label{fig:schema}
\end{figure}

The recently emerging HIN~\cite{shi2017survey} is a good way to model complex relations among different types and objects in recommender systems. Particularly, HIN is a special kind of information network, which either contains multiple types of objects or multiple types of links. The network schema of a HIN specifies the type constraints on the sets of objects and relations among the objects. Two examples used in our experiments are shown in Figure~\ref{fig:schema}. 
In addition, meta-path~\cite{sun2011pathsim}, a relation sequence connecting objects, can effectively extract features of objects and embody rich semantics. In Figure~\ref{fig:amazon}, the meta-path $User$-$Item$-$User$ ($UIU$) extracts the features of users in the purchase history aspect, which means users having the same purchase records. While the $User$-$Item$-$Brand$-$Item$-$User$ ($UIBIU$) extracts the features of users in the brand aspect, which means users purchase the items with the same brand. In the following section, we use the abbreviation to represent the meta-paths. HIN has been widely used in many data mining tasks~\cite{shi2017survey}. HIN based recommendations also have been proposed to utilize rich heterogeneous information in recommender systems, while they usually focus on rating prediction with the ``shallow" model~\cite{shi2016integrating,zhao2017meta}. 

\section{The Proposed Model}
\label{sec:model}
\subsection{Model Framework} 
\label{subsec:framework}

The basic idea of NeuACF is to extract different aspect-level latent features for users and items, and then learn and fuse these latent factors with deep neural network. 
The model contains three major steps. 
First, we construct an HIN based on the rich user-item interaction information in recommender systems, and compute the aspect-level similarity matrices under different meta-paths of HIN which reflect different aspect-level features of users and items. Next, a deep neural network is designed to learn the aspect-level latent factors separately by taking these similarity matrices as inputs. Finally, the aspect-level latent factors are combined with an attention component to obtain the overall latent factors for users and items. Moreover, we also employ self-attention mechanism to fuse aspect-level latent factors more effectively. Next we will elaborate the three steps in the following subsections.

\subsection{Aspect-level Similarity Matrix Extraction}

We employ HIN to organize objects and relations in recommender systems, due to its power of information fusion and semantics representation~\cite{shi2015semantic}. Furthermore, we utilize meta-path to extract different-aspect features of users and items. Taking Figure~\ref{fig:amazon} as an example, we can use $UIU$ and $IUI$ paths to extract features of users and items on the aspect of purchase history, which is extensively exploited by existing latent factor models. In addition, we can also extract features from other aspects. For example, the features of the brand aspect can be extracted from $UIBIU$ and $IBI$ paths . Table \ref{tab:metapath} shows more aspect examples in our experimental datasets.

Given a specific meta-path, there are several alternatives to extract the aspect-level features: commuting matrix or similarity matrix. In this paper, we employ the similarity matrix based on the following reasons. (1) Similarity measure can alleviate noisy information; (2) Similar values within the [0,1] range are more suitable for learning latent factors; (3) Many path based similarity measures are available. We employ the popular 
PathSim~\cite{sun2011pathsim} to calculate aspect-level similarity matrices under different meta-paths in experiments. For example, we compute the similarity matrices of user-user and item-item based on the meta-path $UIBIU$ and $IBI$ for the brand-aspect features.

The computation of similarity matrix based on meta path is of great importance in our propose model, so how to compute similarity matrix quickly is an important problem in our method. In real-word application, the complexity of similarity matrix computation is not high because the similarity matrix is usually very sparse for most meta paths. Based on this fact, there are several acceleration computation methods proposed by previous works~\cite{sun2011pathsim,shi2014hetesim} for similarity matrix computation, for example, PathSim-pruning~\cite{sun2011pathsim}, dynamic programming strategy and Monte Carlo (MC) strategy~\cite{shi2014hetesim}. Moreover there also many new methods for similarity matrix computation, for example, BLPMC~\cite{wei2019prsim}, PRSim~\cite{wang2019accelerating}. In addition, the similarity matrix can be computed offline in advance in our model. The similarity matrix is computed with training data, so we can prepare the similarity matrix before the training processing.

\begin{table}[htb]
  \small 
  \centering 
  \caption{Meta-paths used in experiments and the corresponding aspects.} 
  \label{tab:metapath}
  \begin{tabular}{l|c|c|c} 
  \toprule[\heavyrulewidth]
 \multirow{2}{*}{\textbf{Datasets}} & \multirow{2}{*}{\textbf{Aspect}}& \multicolumn{2}{c}{\textbf{Meta-Paths}} \\
  									\cline{3-4}
  									&&\textbf{User} &\textbf{Movie/Item}  \\
  \hline
  \multirow{3}{*}{MovieLens}  &History    	 	&$UMU$ 	&$MUM$	   \\
							\cline{2-4}
                          	&Director 	    	&$UMDMU$ 	&$MDM$	 \\
							\cline{2-4}
                          &Actor 	    		&$UMAMU$ 	&$MAM$ 			 \\

  \hline
  \multirow{3}{*}{Amazon}  &History    	 	&$UIU$ 	&$IUI$	   \\
							\cline{2-4}
                          	&Brand 	    	&$UIBIU$ 	&$IBI$	 \\
							\cline{2-4}
                          	&Category 	    		&$UICIU$ 	&$ICI$ 			 \\
                            \cline{2-4}
                          	&Co\_view 	    		&$UIVIU$ 	&$IVI$ 			 \\

  \bottomrule[\heavyrulewidth] 
  \end{tabular}
\end{table}

\subsection{Learning Aspect-level Latent Factors}
\begin{figure}
  \centering
  \includegraphics[ width=3.4in]{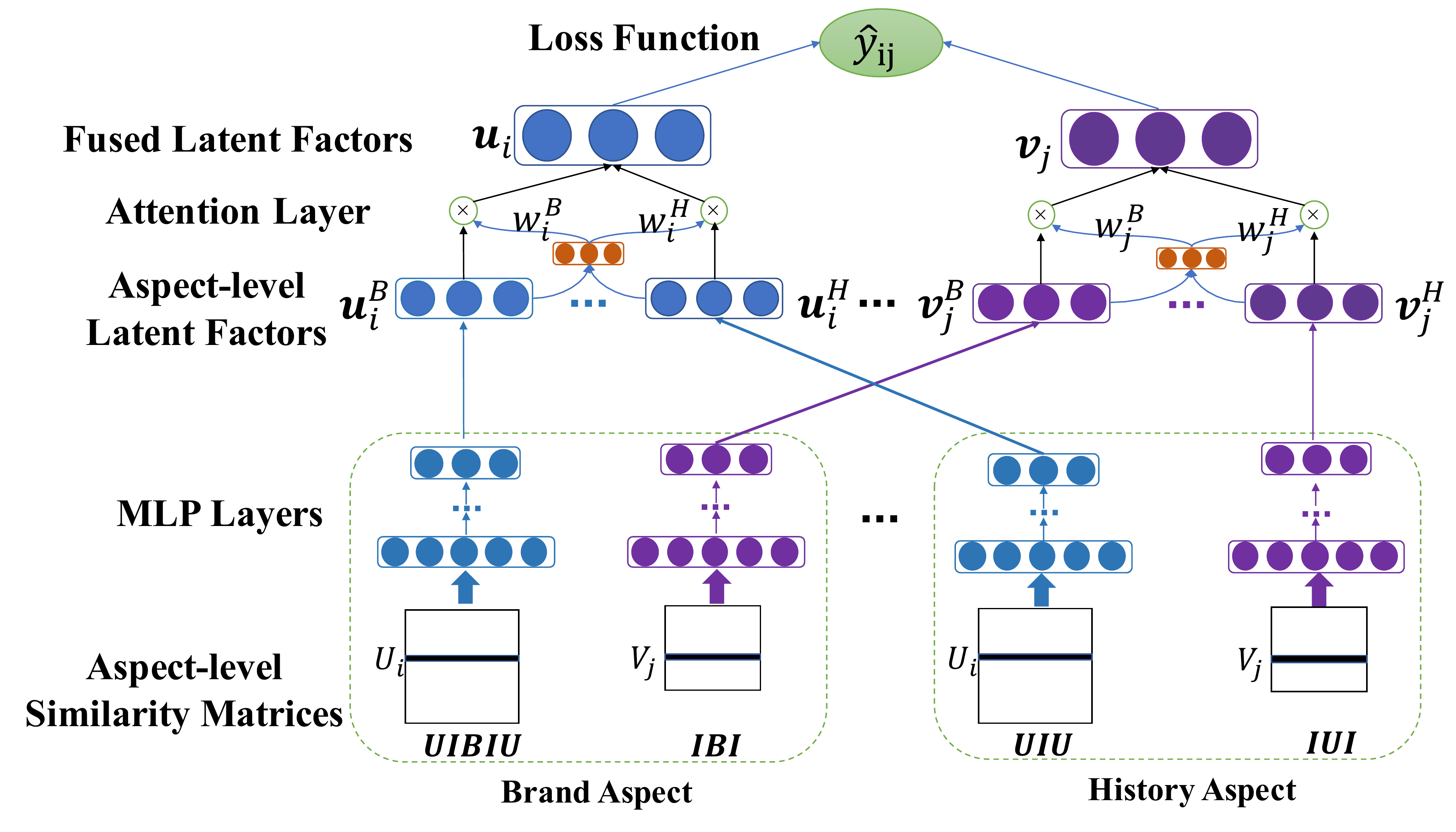}
  \caption{Deep neural network in the NeuACF model.}
  \label{fig:model}
\end{figure}

With the computed user-user and item-item similarity matrices of different aspects, we next learn their latent factors. Different from previous HIN based recommendation models, we design a deep neural network to learn their corresponding aspect-level latent factors separately, and the model architecture is shown in Figure \ref{fig:model}. Concretely, for each user in each aspect, we extract the user's similarity vector from the aspect-specific similarity matrix. Then we take the similarity matrix as the input of the Multi-Layer Perceptron (MLP) and MLP learns the aspect-level latent factor as the output. The item latent factors of each aspect can be learned in a similar way. 
Taking the similarity matrix $\bm{S}^{B}\in \mathbb{R}^{N\times N}$ of users under the meta-path $UIBIU$ as an example, User $U_i$ is represented as an $N$-dimensional vector $\bm{S}^{B}_{i*}$, which means the similarities between $U_i$ and all the other users. Here $N$ means the total number of users in the dataset. 
The MLP projects the initial similarity vector $\bm{S}^{B}_{i*}$ of user $U_i$ to a low-dimensional aspect-level latent factor. In each layer of MLP, the input vector is mapped into another vector in a new space. Formally, given the initial input vector $\bm{S}^{B}_{i*}$, and the $l$-th hidden layer $\bm{H}_{l}$ , the final aspect-level latent factor $\bm{u}^{B}_{i}$ can be learned through the following multi-layer mapping functions,% as Equation~\ref{eq:mlp}.
\begin{equation}
  \label{eq:mlp}
  \begin{split}
  & \bm{H_0} = \bm{S}^{B}_{i*},  \\
  & \bm{H_1} = f( \bm{W}_{1}^{T} * \bm{H}_0 + \bm{b}_1  ), \\
  &\dots\\
  &\bm{H_l} = f( \bm{W}_{l}^{T} * \bm{H}_{l-1} + \bm{b}_{l}  ), \\
  &\dots\\
  &\bm{u}^{B}_{i}= f( \bm{W}_{n}^{T} * \bm{H}_{n-1} + \bm{b}_{n}  ),
  \end{split}
\end{equation}
where $\bm{W}_{i}$ and $\bm{b}_{i}$ are the weight matrix and bias for the $i$-th layer, respectively, and we use the $ReLU$, i.e., $f(x) = max(0,x)$  as the activation function in the hidden layers. 

From the learning framework in Figure~\ref{fig:model}, one can see that for each aspect-level similarity matrix of both users and items there is a corresponding MLP learning component described above to learn the aspect-level latent factors.
As illustrated in Table~\ref{tab:metapath}, for each aspect-level meta-path we can get a corresponding user-user similarity matrix and an item-item similarity matrix. Taking the datasets Amazon as example, we can learn the brand latent factors of users as $\bm{u}^B_i$ and the the brand latent factors of items as $\bm{v}^B_j$ from the meta-path $UIBIU$-$IBI$. Similarly, we can get $\bm{u}^I_i$ and $\bm{v}_j^I$ from the meta-path $UIU$-$IUI$, $\bm{u}_i^C$ and $\bm{v}_j^C$ from the meta-path $UICIU$-$ICI$, as well as $\bm{u}_i^V$ and $\bm{v}_j^V$ from the meta-path $UIVIU$-$IVI$. Since there are variety meta-paths connecting users and items, we can learning different aspect-level latent factors.

\subsection{Attention based Aspect-level Latent Factors Fusion }
\label{sec:attention}
After the aspect-level latent factors are learned separately for users and items, next we need to integrate them together to obtain aggregated latent factors. A straightforward way is to concatenate all the aspect-level latent factors to form a higher-dimensional vector.  Another intuitive way is to average all the latent factors. The issue is that both methods do not distinguish their different importance because not all the aspects contribute to the recommendation equally (we will show that in the experiment part). 
Therefore, we choose the attention mechanism to fuse these aspect-level latent factors. Attention mechanism has shown the effectiveness in various machine learning tasks such as image captioning and machine translation~\cite{xu2015show,you2016image,bahdanau2014neural}. The advantage of attention mechanism is that it can learn to assign attentive values (normalized by sum to 1) for all the aspect-level latent factors: higher (lower) values indicate that the corresponding features are more informative (less informative) for recommendation.
Specifically, given the user's brand-aspect latent factor $\bm{u}^{B}_{i}$, we use a two-layers network to compute the attention score $\bm{s}_i^B$ by the following Equation~\ref{equ:atteion-s},
\begin{equation}
  \label{equ:atteion-s}
  \bm{s}_i^B = \bm{W}_2^{T}f\left( \bm{W}_{1}^T* \bm{u}^{B}_{i} + \bm{b_1} \right) + \bm{b_2},
\end{equation}
where  $\bm{W_*}$ is the weight matrices and $\bm{b_*}$ is the biases.

The final attention values for the aspect-level latent factors are obtained by normalizing the above attentive scores with the Softmax function given in Equation~\ref{equ:atteion-softamx}, which can be interpreted as the contributions of different aspects $B$ to the aggregated latent factor of user $U_i$,
\begin{equation}
  \label{equ:atteion-softamx}
  \bm{w}_i^B = \frac{ exp( \bm{s}_i^B ) }{ \sum_{A=1}^{L} exp( \bm{s}_i^A ) }, 
\end{equation}
where $L$ is the total number of all the aspects.

After obtaining all the attention weights $\bm{w}_i^B$ of all the aspect-level latent factors for user $U_i$, the aggregated latent factor $\bm{u}_{i}$ can be calculated by Equation~\ref{equ:atteion-sum},
\begin{equation}
  \label{equ:atteion-sum}
  \bm{u}_{i} = \sum_{B=1}^{L} \bm{w}_i^B \centerdot \bm{u}^{B}_{i}.
\end{equation}

We implement this attention method as NeuACF in our experiments.

\subsection{Self-Attention based Aspect-level Latent Factors Fusion }
\label{sec:self_attention}
% self attention co-attention and related work
Recently, self-attention mechanism has received considerable research interests. 
For example, Vaswani et al.~\cite{vaswani2017attention} and Devlin et al.~\cite{devlin2018bert} utilize self-attention to learn the relationship between two sequences. 
Learning dependencies and relationships between  aspect-level latent factors is the most important part in our model, and self-attention has ability to model the relationships between the different aspect-level latent factors.

Different from standard attention mechanism, self-attention mainly focuses on the co-learning attentions of two sequences. 
The vanilla attention mechanism mainly considers computing the attention values based on the user or item representations of one aspect, while self-attention mechanism is able to learn the attention values from different aspects simultaneously. For example, the Brand-level latent factor of users have strong relationship to the  Brand-level latent factor of items, and the self-attention mechanism can learn this relationship and promote the performance of recommendation. So the learned values are able to capture more information on the multi-aspects.
In details, we firstly compute the affinity scores between all aspect-level latent factors. For a user $U_i$, the affinity score of two different aspect-level latent factors $\bm{u}_i^B$ and $\bm{u}_i^C$ can be calculated by their inner product:
\begin{equation}
M_{i}^{B,C} = {(\bm{u}_{i}^{B})}^T*\bm{u}_{i}^{C}.
\end{equation}

The matrix $\bm{M}_i=[M_{i}^{B,C}] \in \mathbb{R}^{L\times L} $ is also called the self-attention matrix, where $L$ is the total number of aspects. In fact, there is an affinity matrix $\bm{M}_i$ for each user. Basically, the matrix $\bm{M_i}$ characterizes the similarity of aspect-level latent factors for the specific user $U_i$, which reflects the correlation between two aspects when recommending for this user. When the aspect $B$ is equal to aspect $C$, $M_{i}^{B,C}$ will get a high value due to the inner product operator, so we add a zero mask to avoid a high matching score between identical vectors. 

The aspect-level latent factors learned from self-attention mechanism are not independent. Users will make a trade-off between those aspects. The affinity matrix measures the importance of different aspect-level latent factors, so we compute the representation of aspect $B$ for the specific user $i$ based on the self-attention matrix as:
\begin{equation}
\label{equ:att1}
\bm{g}_i^B = \sum_{C=1}^{L}\frac{exp(\bm{M}_{i}^{B,C})}{\sum_{A=1}^{L}exp(\bm{M}_{i}^{B,A})}\bm{u}_{i}^{C}.
\end{equation}

Then for all the aspects, we can obtain the final representation of users or items as:
\begin{equation}
\label{equ:att2}
\bm{u}_i = \sum_{B=1}^{L}\bm{g}_i^B.
\end{equation}

The self-attention mechanism can learn self-attentive representations from different aspect-level information effectively. 
In order to distinguish with the above attention method NeuACF, we implement the self-attention mechanism as NeuACF++ in our experiments.

\subsection{Objective Function}
\label{subsec:optimization}
We model the top-N recommendation as a classification problem which predicts the probability of interaction between users and items in the future. In order to ensure that the output value is a probability, we need to constrain the output $\hat{y}_{ij}$ in the range of [0,1], where we use a Logistic function as the activation function for the output layer. The probability of the interaction between the user $U_i$ and item $I_j$ is calculated according to Equation~\ref{equ:last},
\begin{equation}
  \label{equ:last}
  \hat{y}_{ij} = sigmod( \bm{u}_i*\bm{v}_j ) = \frac{1}{1+e^{-\bm{u}_i*\bm{v}_j}},
\end{equation}
where $\bm{u}_i$ and $\bm{v}_j$ are the aggregated latent factors of user $U_i$ and item $I_j$ respectively.

Over all the training set, according to the above settings, the likelihood function is:
\begin{equation}
  \label{equ:likelihood}
  p( \mathcal{Y},\mathcal{Y}^{-}|\Theta ) = \prod_{i,j\in\mathcal{Y}} \hat{y}_{ij}\prod_{i,k\in\mathcal{Y}^{-}} (1- \hat{y}_{ik} ),
\end{equation}
where $\mathcal{Y}$ and $\mathcal{Y}^{-}$ are the positive and negative instances sets, respectively. The negative instance set $\mathcal{Y}^{-}$ is sampled from unobserved data for training. $\Theta$ is the parameters set.

Since the ground truth $y_{ij}$ is in the set $\{0, 1\}$, Equation~\ref{equ:likelihood} can be rewritten as:
\begin{equation}
    p( \mathcal{Y},\mathcal{Y}^{-}|\Theta ) = \prod_{i,j\in\mathcal{Y} \cup \mathcal{Y}^{-}} (\hat{y}_{ij})^{y_{ij}} * (1- \hat{y}_{ij} )^{(1 - y_{ij})}.
\end{equation}

Then we take the negative logarithm of the likelihood function to get the point-wise loss function in Equation~\ref{equ:loss},
\begin{equation}
    \label{equ:loss}
    Loss = -\sum_{i,j\in \mathcal{Y} \cup\mathcal{Y}^{-} }  \left( y_{ij}log\hat{y}_{ij} + (1-y_{ij})log(1-\hat{y}_{ij}) \right),
\end{equation}
where $y_{ij}$ is the ground truth of the instance and $\hat{y}_{ij}$ is predicted score. This is the overall objective function of our model, and we can optimize it by stochastic gradient descent or its variants~\cite{kingma2014adam}.

\subsection{ Discussion }
Here, we give the analysis of our proposed models NeuACF and NeuACF++. 

\begin{itemize}

\item NeuACF and NeuACF++ are general frameworks for recommendation. We can learn aspect-level latent factors from aspect-level features computed via different methods. For example, the similarity matrix $\bm{S}^B$ can also be computed with HeteSim~\cite{shi2014hetesim} or PCRW~\cite{lao2010relational}.

\item As a deep neural network model, DMF~\cite{zhang2017deep} can be considered as one special case of our model. 
DMF does not take the heterogeneous information into consideration, so 
if we only consider the user-item purchase history aspect, our model is equivalent to the DMF model. We argue that the aspect information learned from meta-paths has potential to increase the performance of recommendation.

\item We present the time complexity analysis of our proposed models NeuACF and NeuACF++ here. Generally, the time complexity is affected by the epochs of iterator $T$, the size of training sample $S$, the number of aspects $L$ and the size of hidden numbers $H$. When we utilize three-layer MLP to learn user and item latent factors in our models, the time complexity of forward and backward process is bounded by matrix multiplication. Let $h_{n_1}$ be the number of input neurons and $h_{n_2}$ be the number of output neurons, the time complexity of forward process can be calculated as $O(h_{n_1}*H + H*h_{n_2})$. The attention layer is a two-layer neural network with the number of input size equal to $h_{n_2}$ and the number of output size is 1. The time consumption is negligible comparing to the embedding layers. Therefore, the overall time complexity for training process is $O(STL(h_{n_1}*H + H*h_{n_2}))$. For the prediction process, supposing the number of negative sampling for one user is $N_s$, the time complexity of prediction is $O(N_sL(h_{n_1}*H + H*h_{n_2}))$.
\end{itemize}

\section{Experiments}
\label{sec:experiment}
\subsection{Experimental Settings}
\subsubsection{Datasets}

We evaluate the proposed model over the publicly available MovieLens dataset~\cite{harper2016movielens} and Amazon dataset~\cite{he2016ups,mcauley2015image}. We use the origin Movielens dataset for our experiment. For Amazon dataset, we remove the users who buy less than 10 items. The network schema is shown in Figure~\ref{fig:schema}, and the statistics of the datasets are summarized in Table ~\ref{tab:statistics}.

\begin{itemize}
  \item  MovieLens-100K(ML100k)/MovieLens-1M(ML1M) \footnote{https://grouplens.org/datasets/movielens/}: MovieLens datasets have been widely used for movie recommendation. We use the versions ML100K and ML1M. For each movie, we crawl the directors, actors of the movie from IMDb. 
  \item Amazon\footnote{http://jmcauley.ucsd.edu/data/amazon/}: This dataset contains users' rating data in Amazon. In our experiment, we select the items of Electronics categories for evaluation.
\end{itemize}

\begin{table}
  \small
  \centering
  \caption{The statistics of the datasets.}
  \label{tab:statistics}
  \begin{tabular}{c|c|c|c|c}
  \hline
  \textbf{Dataset}      &\textbf{\#users}  &\textbf{\#items}  &\textbf{\#ratings}     &\textbf{\#density}  \\\hline
  ML100K               &943      &1682     &100,000       &6.304\%  \\ \hline
  ML1M                 &6040     &3706     &1,000,209     &4.468\%  \\ \hline
  Amazon                &3532     &3105     &57,104        &0.521\%   \\ \hline
  \end{tabular}
\end{table}
\subsubsection{Evaluation Metric}
We adopt the leave-one-out method~\cite{he2017neural,xue2017deep} for evaluation. The latest rated item of each user is held out for testing, and the remaining data for training. Following previous works~\cite{he2017neural,xue2017deep}, we randomly select 99 items that are not rated by the users as negative samples and rank the 100 sampled items for the users. For a fair comparison with the baseline methods, we use the same negative sample set for each (\emph{user, item}) pair in the test set for all the methods. We evaluate the model performance through the Hit Ratio (HR) and the Normalized Discounted Cumulative Gain (NDCG) defined in Equation~\ref{equ:hrndcg},
\begin{equation}
  \label{equ:hrndcg}
  HR = \frac{\#hits}{\#users},NDCG = \frac{1}{\#users}\sum_{i=1}^{\#users}\frac{1}{log_{2}(p_i+1)},
\end{equation}
where $\#hits$ is the number of users whose test item appears in the recommended list and $p_i$ is the position of the test item in the list for the $i$-th hit. In our experiments, we truncate the ranked list at $K\in[5,10,15,20]$ for both metrics. 

\subsubsection{Baselines}
Besides two basic methods (i.e., ItemPop and ItemKNNN~\cite{sarwar2001item}), the baselines include two MF methods (MF~\cite{koren2009matrix} and eALS~\cite{he2016fast}), one pairwise ranking method (BPR~\cite{rendle2009bpr}), and two neural network based methods (DMF~\cite{xue2017deep} and NeuMF~\cite{he2017neural}). In addition, we use $\text{SVD}_{\text{hin}}$ to leverage the heterogeneous information for recomendation, and we also adopt two recent HIN based methods (FMG\\ \cite{zhao2017meta} and HeteRs~\cite{pham2016general}) as baselines.
\begin{itemize}
 \item ItemPop. Items are simply ranked by their popularity judged by the number of interactions. This is a widely-used non-personalized method to benchmark the recommendation performance.
\item ItemKNN~\cite{sarwar2001item}. It is a standard item-based collaborative filtering method. 
\item MF~\cite{koren2009matrix}. Matrix factorization is a representative latent factor model. 
\item eALS~\cite{he2016fast}. It is a state-of-the-art MF method for recommendation with the square loss. 
\item BPR~\cite{rendle2009bpr}. The Bayesian Personalized Ranking approach optimizes the MF model with a pairwise ranking loss, which is tailored to learn from implicit feedback.
\item DMF~\cite{xue2017deep}. DMF uses the interaction matrix as the input and maps users and items into a common low-dimensional space using a deep neural network. %We implemented the model based on the paper.
\item NeuMF~\cite{he2017neural}. It combines the linearity of MF and non-linearity of DNNs for modelling user–item latent structures. In our experiments, we use the NeuMF with pre-trained, We used hyper-parameters followed the instructions in the paper.
\item  $\text{SVD}_{\text{hin}}$. SVDFeature~\cite{chen2012svdfeature} is designed to efficiently solve the feature-based matrix factorization. $\text{SVD}_{\text{hin}}$ uses SVDFeature to leverage the heterogeneous information for recommendation. Specifically, we extract the heterogeneous information (e.g.attributes of movies/items and profiles of users) as the input of SVDFeature.
\item HeteRS~\cite{pham2016general}. HeteRS is a graph-based model which can solve general recommendation problem on heterogeneous networks. It models the rich information with a heterogeneous graph and considers the recommendation problem as a query-dependent node proximity problem.
\item FMG~\cite{zhao2017meta}. It proposes ``MF+FM'' framework for the HIN-based rating prediction. We modify its optimization object as point-wise ranking loss for the top-N recommendation. 
\end{itemize}  

\subsubsection{Implementation}
We implement the proposed NeuACF and NeuACF++ based on Tensorflow~\cite{abadi2016tensorflow}. We use the same hyper-parameters for all the datasets. 
For the neural network, we use a three-layer MLP with each hidden layer having 600 hidden units. The dimension of latent factors is 64. We randomly initialize the model parameters with a xavier initializer~\cite{glorot2010understanding}, and use the Adam ~\cite{kingma2014adam} as the optimizer. We set the batch size to 1024 and set the learning rate to 0.0005. When training our model, 10 negative instances are sampled for each positive instance. Table \ref{tab:metapath} illustrates the extracted aspects and corresponding meta-paths. Some meta-paths are also used for FMG. 
The optimal parameters for baselines are set according to literatures. 
All the experiments are conducted on a machine with two GPUs (NVIDIA GTX-1080 *2) and two CPUs (Intel Xeon E5-2690 * 2).

\subsection{Experiment Results}

\begin{table*}
  \scriptsize
  \centering 
 
   \caption{HR@K and NDCG@K comparisons of different methods.}
  \label{tab:performance}
  \begin{tabular}{l|c||c|c|c|c|c|c|c|c|c|c|c|c} 
  \toprule[\heavyrulewidth]
  
  \textbf{Datasets} &\textbf{Metrics}& \textbf{ItemPop}& \textbf{ItemKNN} & \textbf{MF} & \textbf{eALS}&  \textbf{BPR}& \textbf{DMF}& \textbf{NeuMF} &\textbf{$\text{SVD}_{\text{hin}}$} &\textbf{HeteRS} &\textbf{FMG}  &\textbf{NeuACF} & \textbf{NeuACF++} \\ 
\hline
\hline
\multirow{8}{*}{ML100K}   & HR@5    &0.2831   &0.4072   &0.4634   &0.4698   &0.4984           &0.3483   &0.4942   &0.4655   &0.3747    &0.4602     &0.5097  &\textbf{0.5111}	\\
                          & NDCG@5  &0.1892   &0.2667   &0.3021   &0.3201   &0.3315           &0.2287   &0.3357   &0.3012   &0.2831    &0.3014     &0.3505  &\textbf{0.3519} 	\\
                          \cline{2-14}
                          & HR@10   &0.3998   &0.5891   &0.6437   &0.6638   &0.6914           &0.4994   &0.6766   &0.6554   &0.5337     &0.6373      &0.6846   &\textbf{0.6915}	  	\\
                          & NDCG@10 &0.2264   &0.3283   &0.3605   &0.3819   &0.3933           &0.2769   &0.3945   &0.3988   &0.3338     &0.3588     &0.4068   &\textbf{0.4092}   	\\
                          \cline{2-14}
                          & HR@15   &0.5366   &0.7094   &0.7338   &0.7529   &0.7741           &0.5873   &0.7635   &0.7432   &0.6524     &0.7338     &0.7813  &\textbf{0.7832}    	\\
                          & NDCG@15 &0.2624   &0.3576   &0.3843   &0.4056   &0.4149           &0.3002   &0.4175   &0.4043   &0.3652     &0.3844     &0.4318  &\textbf{0.4324}    	\\
                          \cline{2-14}
                          & HR@20   &0.6225   &0.7656   &0.8144   &0.8155   &0.8388           &0.6519   &0.8324   &0.8043   &0.7224     &0.8006     &\textbf{0.8464}   &0.8441    	\\
                          & NDCG@20 &0.2826   &0.3708   &0.4034   &0.4204   &0.4302           &0.3151   &0.4338   &0.3944   &0.3818     &0.4002     &\textbf{0.4469}	&\textbf{0.4469}	\\
  
\hline
\hline                        
\multirow{8}{*}{ML1M}     & HR@5    &0.3088   &0.4437   &0.5111   &0.5353   &0.5414           &0.4892   &0.5485    &0.4765   &0.3997   &0.4732     &\textbf{0.5630} &	0.5584	\\
                          & NDCG@5  &0.2033   &0.3012   &0.3463   &0.3670   &0.3756           &0.3314   &0.3865    &0.3098   &0.2895   &0.3183     &\textbf{0.3944}	&0.3923	\\
                          \cline{2-14}
                          & HR@10   &0.4553   &0.6171   &0.6896   &0.7055   &0.7161           &0.6652   &0.7177    &0.6456   &0.5758   &0.6528     &0.7202 &\textbf{0.7222}	\\
                          & NDCG@10 &0.2505   &0.3572   &0.4040   &0.4220   &0.4321           &0.3877   &0.4415    &0.3665   &0.3461   &0.3767     &0.4453 &\textbf{0.4454}	\\
                          \cline{2-14}
                          & HR@15   &0.5568   &0.7118   &0.7783   &0.7914   &0.7988           &0.7649   &0.7982    &0.7689   &0.6846   &0.7536     &0.8018 & \textbf{0.8030}\\
                          & NDCG@15 &0.2773   &0.3822   &0.4275   &0.4448   &0.4541           &0.4143   &0.4628    &0.4003   &0.3749   &0.4034     &\textbf{0.4667} & 0.4658\\
                          \cline{2-14}
                          & HR@20   &0.6409   &0.7773   &0.8425   &0.8409   &0.8545           &0.8305   &0.8586    &0.8234   &0.7682   &0.8169     &0.8540 &\textbf{0.8601}	\\
                          & NDCG@20 &0.2971   &0.3977   &0.4427   &0.4565   &0.4673           &0.4296   &0.4771    &0.4456   &0.3947   &0.4184     &0.4789 &\textbf{0.4790}	\\
\hline
\hline                        
\multirow{8}{*}{Amazon}   & HR@5    &0.2412   &0.1897   &0.3027   &0.3063   &0.3296           &0.2693   &0.3117     &0.3055   &0.2766   &0.3216    &0.3268	&\textbf{0.3429}\\
                          & NDCG@5  &0.1642   &0.1279   &0.2068   &0.2049   &0.2254           &0.1848   &0.2141     &0.1922   &0.1800   &0.2168    &0.2232   &\textbf{0.2308}\\
                          \cline{2-14}
                          & HR@10   &0.3576   &0.3126   &0.4278   &0.4287   &0.4657           &0.3715   &0.4309     &0.4123   &0.4207   &0.4539    &0.4686  &\textbf{0.4933}\\
                          & NDCG@10 &0.2016   &0.1672   &0.2471   &0.2441   &0.2693           &0.2179   &0.2524     &0.2346   &0.2267   &0.2595    &0.2683  &\textbf{0.2792}\\
                          \cline{2-14}
                          & HR@15   &0.4408   &0.3901   &0.5054   &0.5065   &0.5467           &0.4328   &0.5258     &0.5056   &0.5136   &0.5430    &0.5591  &\textbf{0.5948}\\
                          & NDCG@15 &0.2236   &0.1877   &0.2676   &0.2647   &0.2908           &0.2332   &0.2774     &0.2768   &0.2513   &0.2831    &0.2924   &\textbf{0.3060}\\
                          \cline{2-14}
                          & HR@20   &0.4997   &0.4431   &0.5680   &0.5702   &0.6141           &0.4850   &0.5897     &0.5607   &0.5852   &0.6076    &0.6257  &\textbf{0.6702}\\
                          & NDCG@20 &0.2375   &0.2002   &0.2824   &0.2797   &0.3067           &0.2458   &0.2925     &0.2876   &0.2683   &0.2983    &0.3080   &\textbf{0.3236}\\

  \bottomrule[\heavyrulewidth] 
  \end{tabular}
\end{table*}

\subsubsection{Performance Analysis}
Table~\ref{tab:performance} shows the experiment results of different methods. Our proposed methods are marked as NeuACF which implements the attention method in Section~\ref{sec:attention} and NeuACF++ which implements the self-attention mechanism in Section~\ref{sec:self_attention}, respectively. One can draw the following conclusions.

Firstly, one can observe that, NeuACF and NeuACF++ achieve all the best performance over all the datasets and criteria. The improvement of the two models comparing to these baselines is significant. This indicates that the aspect level information is useful for recommendations. 
Besides, NeuACF++ outperforms the NeuACF method in most circumstances. Particularly, the performance of NeuACF++ is significantly improved in Amazon dataset about (+2\% at HR and +1\% at NDCG). This demonstrates the effectiveness of the self-attention mechanism. Since the affinity matrix evaluates the similarity score of different aspects, we can extract the valuable information from the aspect latent factors.

Secondly, NeuMF, as one neural network based method, also performs well on most conditions, while both NeuACF and NeuACF++ outperform NeuMF in almost all the cases. The reason is probably that multiple aspects of latent factors learned by NeuACF and NeuACF++ provide more features of users and items. Although FMG also utilizes the same features with NeuACF and NeuACF++, the better performance of NeuACF and NeuACF++ implies that the deep neural network and the attention mechanisms in NeuACF and NeuACF++ may have the better ability to learn latent factors of users and items than the ``shadow'' model in FMG.

We can also observe that MF based methods outperform the ItemPop and ItemKNN methods. This indicates that the latent factors models can depict the user and item characteristics. Moreover, the performance of NeuMF is better than MF, which indicates that the non-linear projection can capture more information. The performance of BPR is comparable to NeuMF though it does not utilize the non-linear projection. The reason may be that the objective function is prone to tackle those ranking problems. 

\subsubsection{Impact of Different Aspect-level Latent Factors}
\label{sec:single_aspect}
\begin{figure}
  \centering
  \begin{subfigure}[ht]{0.23\textwidth}
      \centering
      \includegraphics[height=1.5in]{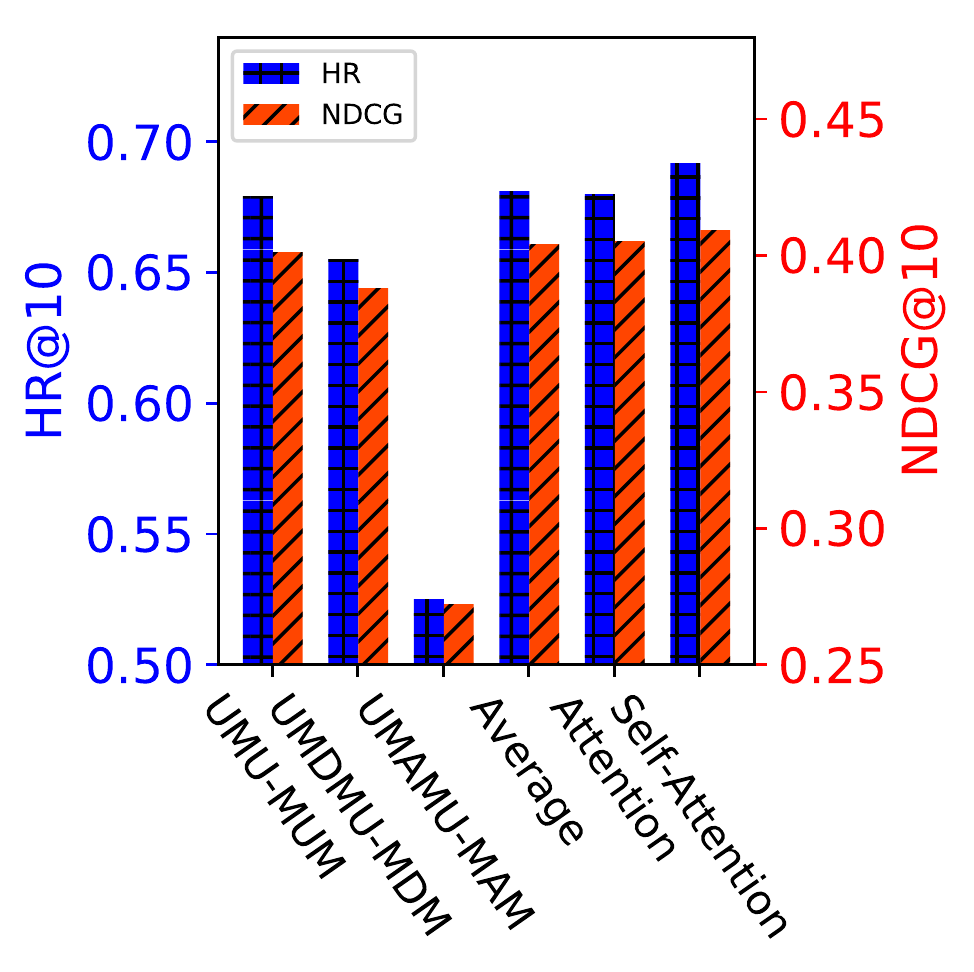}
      \caption{ML100K: single aspect}
      \label{fig:ml-100k_metapath}
  \end{subfigure}
  ~
  \begin{subfigure}[ht]{0.23\textwidth}
      \centering
      \includegraphics[height=1.5in, width=1.5in]{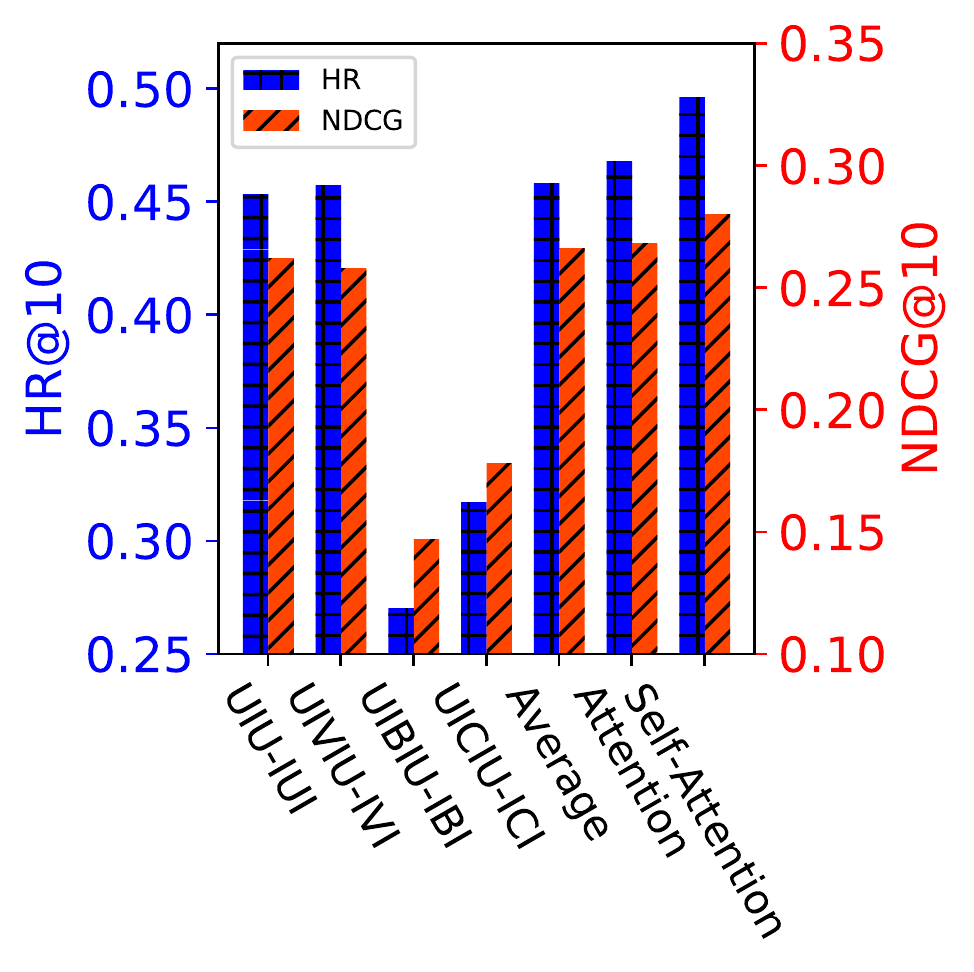}
      \caption{Amazon: single aspect}
      \label{fig:amazon_metapath}
  \end{subfigure}
  ~
  \begin{subfigure}[ht]{0.23\textwidth}
      \centering
      \includegraphics[height=1.4in]{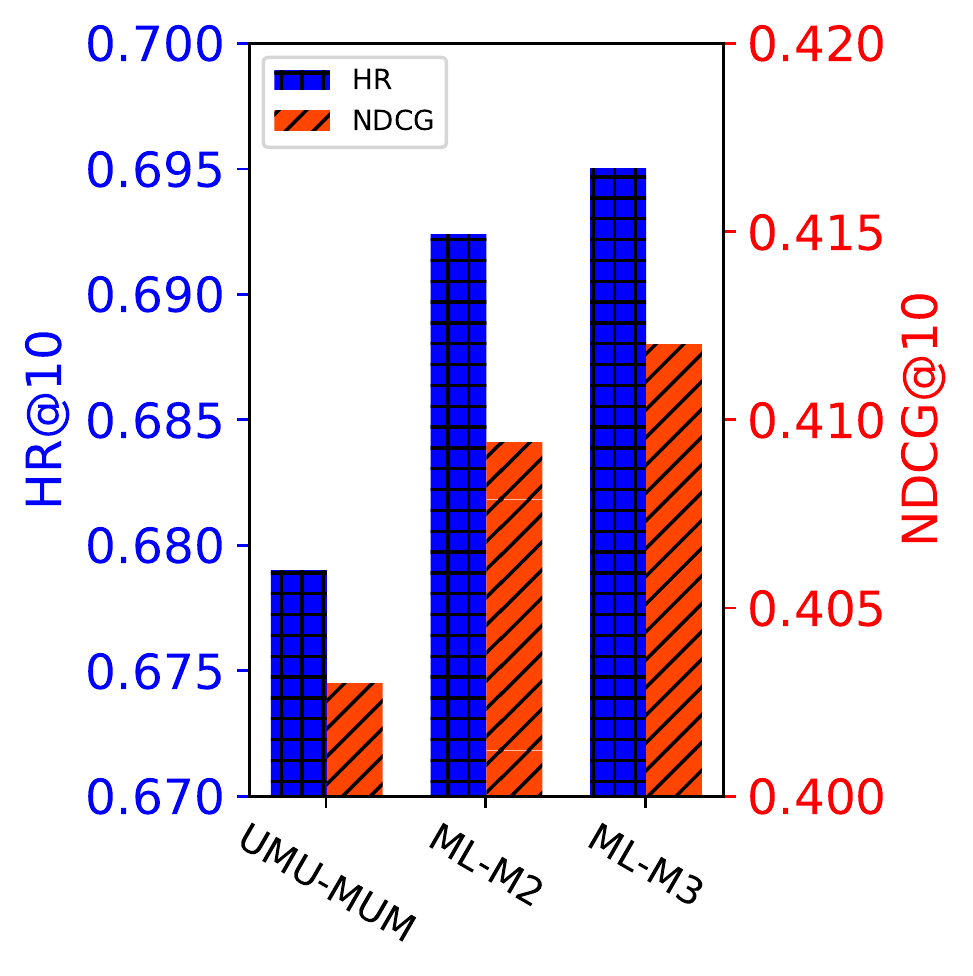}
      \caption{ML-100k: fusing aspects}
      \label{fig:ml-100k_metapath_add}
  \end{subfigure}
  ~
  \begin{subfigure}[ht]{0.23\textwidth}
      \centering
      \includegraphics[height=1.4in]{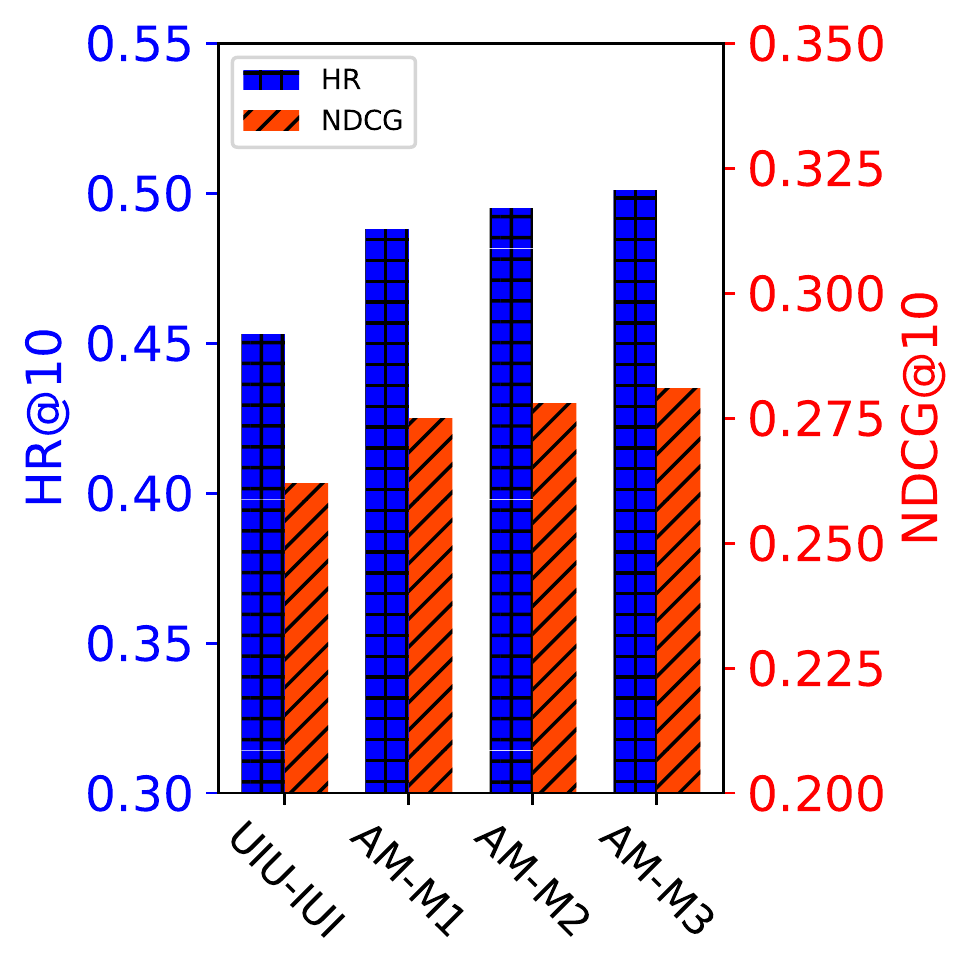}
      \caption{Amazon: fusing aspects}
      \label{fig:amazon_metapath_added}
  \end{subfigure}
  \caption{The impact of different aspect-level latent factors. (a) The performance of single aspect on $MovieLens$. ``Attention'' means the NeuACF method, and ``Self-Attention'' means the NeuACF++ method. (b) The performance of single aspect on Amazon dataset. (c) The performance of combination of different meta-paths on ML100k dataset. ML-M2 adds $UMDMU$-$MDM$, and ML-M3 adds $UMAMU$-$MAM$ to ML-M2.
(d) The performance of combination of different meta-paths on Amazon dataset. AM-M1 adds $UIVIU$-$IVI$. AM-M2 and AM-M3 add $UIBIU$-$IBI$, $UICIU$-$ICI$, respectively.}
  \label{fig:metapath}
\end{figure}

To analyze the impact of different aspect-level latent factors on the algorithm performance, we run NeuACF and NeuACF++ with individual aspect-level latent factor through setting meta-paths. In Figure~\ref{fig:metapath}, for example, $UIBIU$-$IBI$ means that we only learn the brand-aspect latent factor for users and items. In addition, we also run NeuACF with the ``Average'', ``Attention'' and  ``Self-Attention'' fusion mechanisms, where ``Average'' means averaging all the aspect-level latent factors, ``Attention'' means fusing latent factors with the proposed attention mechanism in Section~\ref{sec:attention}, and ``Self-Attention'' means fusing latent factors with the self-attention mechanism mentioned in Section~\ref{sec:self_attention}. From the results shown in Figure~\ref{fig:ml-100k_metapath} and Figure~\ref{fig:amazon_metapath}, one can observe that the purchase-history aspect factors (e.g., $UMU$-$MUM$ and $UIU$-$IUI$) usually get the best performance in all the individual aspects which indicates that the purchase history of users and items usually contains the most important information. 
One can also see that ``Average'', ``Attention'' and ``Self-Attention'' always perform better than individual meta-path, demonstrating fusing all the aspect-level latent factors can improve the performance. In addition, the better performance of ``Attention'' than ``Average'' also shows the benefit of the attention mechanism in NeuACF. One can also observe that the ``Self-Attention'' mechanism always perform better than other methods, which indicates that the self-attention mechanism can fuse different aspect information more efficiently.

Further, in order to validate that the additional information from different meta-paths has potential to increase the recommendation performance. We conduct experiments with the increase of meta-paths to fuse more information into our proposed models. The results are shown in Figure~\ref{fig:ml-100k_metapath_add} and Figure~\ref{fig:amazon_metapath_added}. It demonstrates that the combination of different meta-paths can increase the performance of recommendation. In particular, ML-M2 means the result of fusing aspect-level latent factors extracted from the meta-paths of $UMU$-$MUM$ and $UMAMU$-$MAM$. The performance of ML-M2 outperforms the single meta-path $UMU$-$MUM$, which is the best result among all the single aspects. ML-M3 means the result of fusing the meta-paths of $UMU$-$MUM$, $UMAMU$-$MAM$ and $UMDMU$-$MDM$. Similarly, the result is better than ML-M2. 
Moreover, the performance does not improve linearly. Taking the Amazon dataset in Figure~\ref{fig:amazon_metapath_added} as an example, the meta-path $UIVIU$-$IVI$ in AM-M1, comparing to the single meta-path $UIU$-$IUI$, provides a large improvement. However, the meta-path $UIBIU$-$IBI$ in AM-M2 helps little on the performance. This demonstrates that different aspect-level meta-paths contain unequal information, so it is essential to automatically fuse aspect-level latent factors with attention mechanisms.

 \begin{figure}
  \centering
  \begin{subfigure}[ht]{0.25\textwidth}
      \centering
      \includegraphics[height=1.5in, width=1.6in]{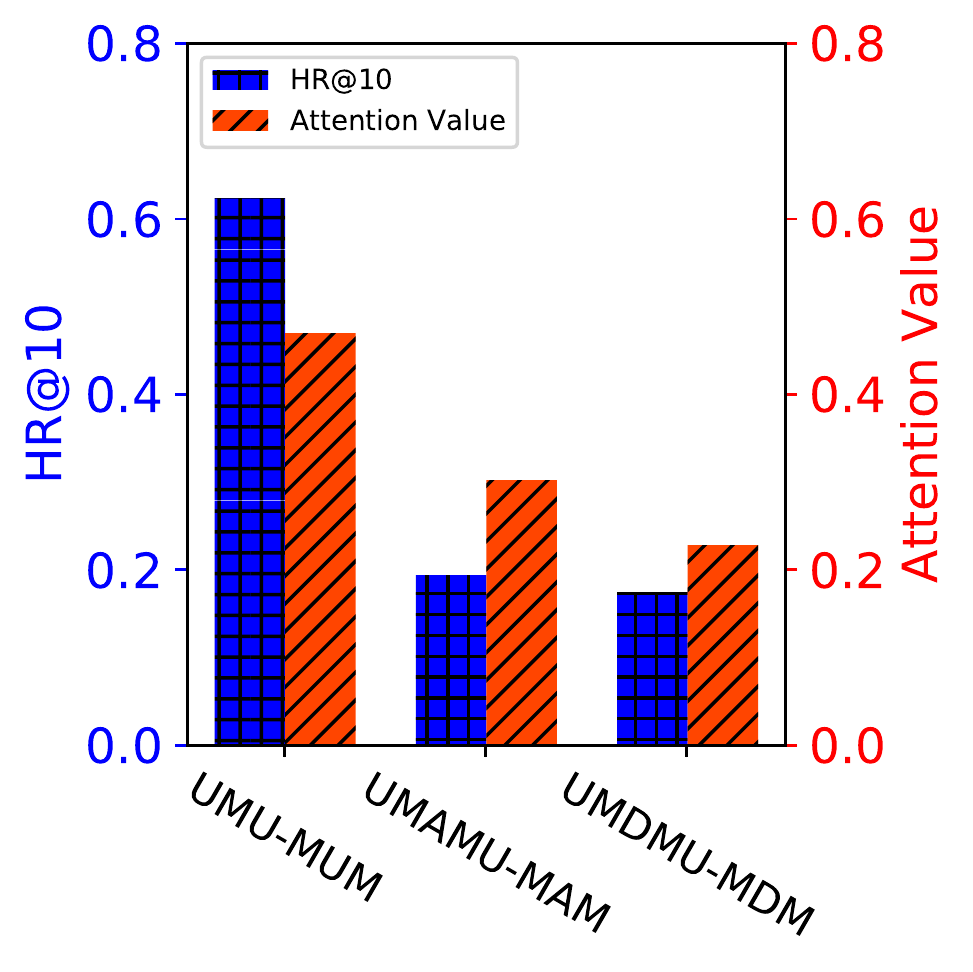}
      \caption{ML100k: NeuACF}
      \label{fig:ml-att_NeuACF}
  \end{subfigure}~
  \centering
  \begin{subfigure}[ht]{0.25\textwidth}
      \centering
      \includegraphics[height=1.5in, width=1.6in]{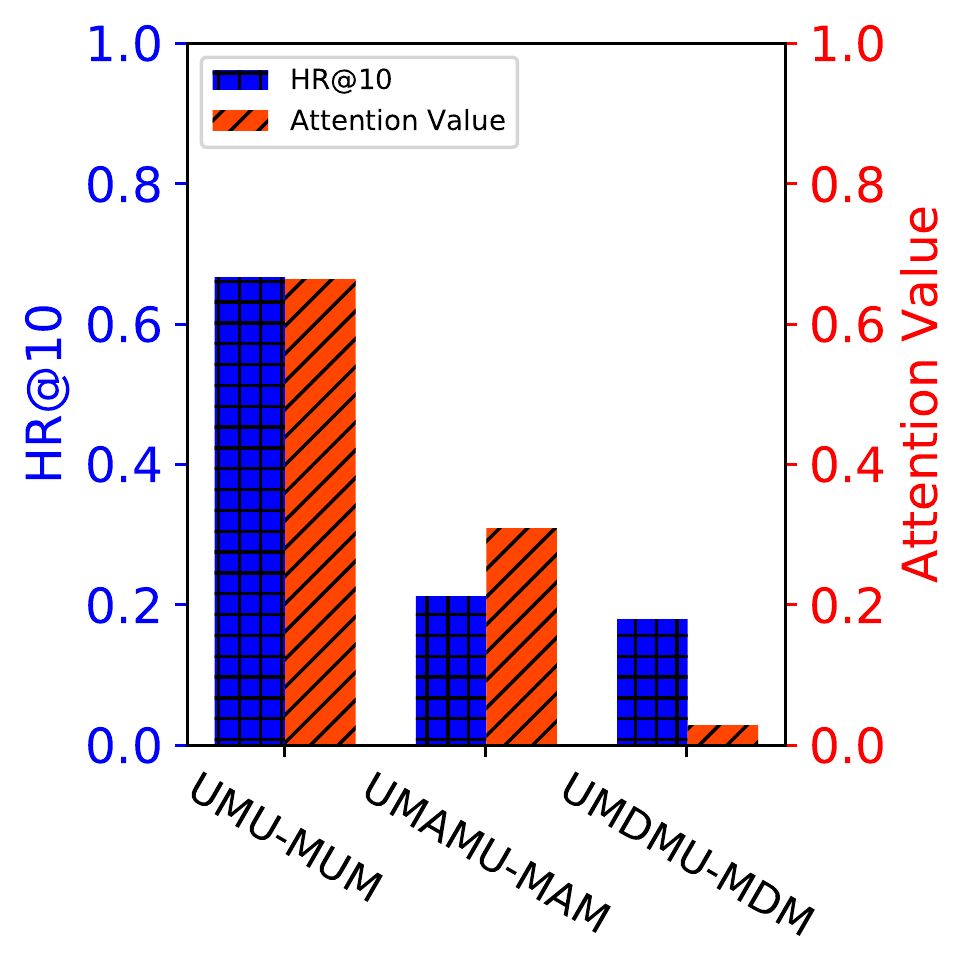}
      \caption{ML100k: NeuACF++}
      \label{fig:ml-att}
  \end{subfigure}
  \centering
  \begin{subfigure}[ht]{0.25\textwidth}
      \centering
      \includegraphics[height=1.5in, width=1.6in]{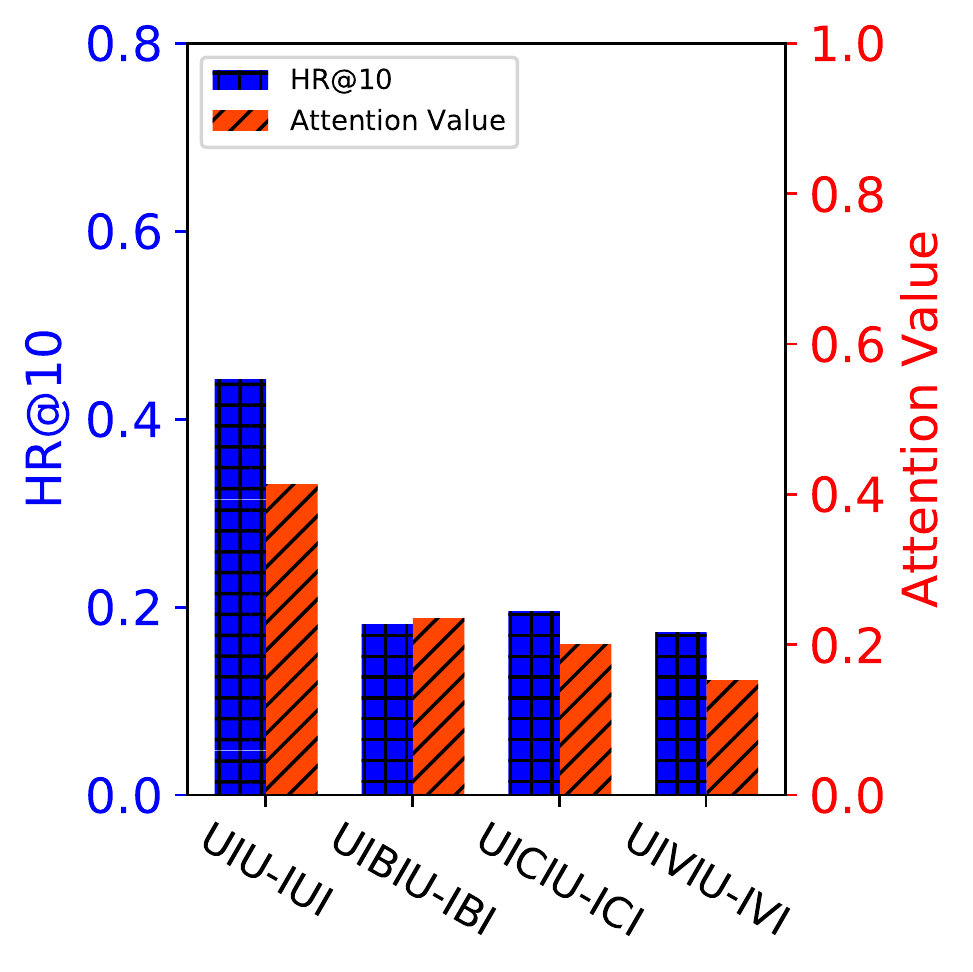}
      \caption{Amazon: NeuACF}
      \label{fig:amazon_att_NeuACF}
  \end{subfigure}~
  \centering
  \begin{subfigure}[ht]{0.25\textwidth}
      \centering
      \includegraphics[height=1.5in, width=1.6in]{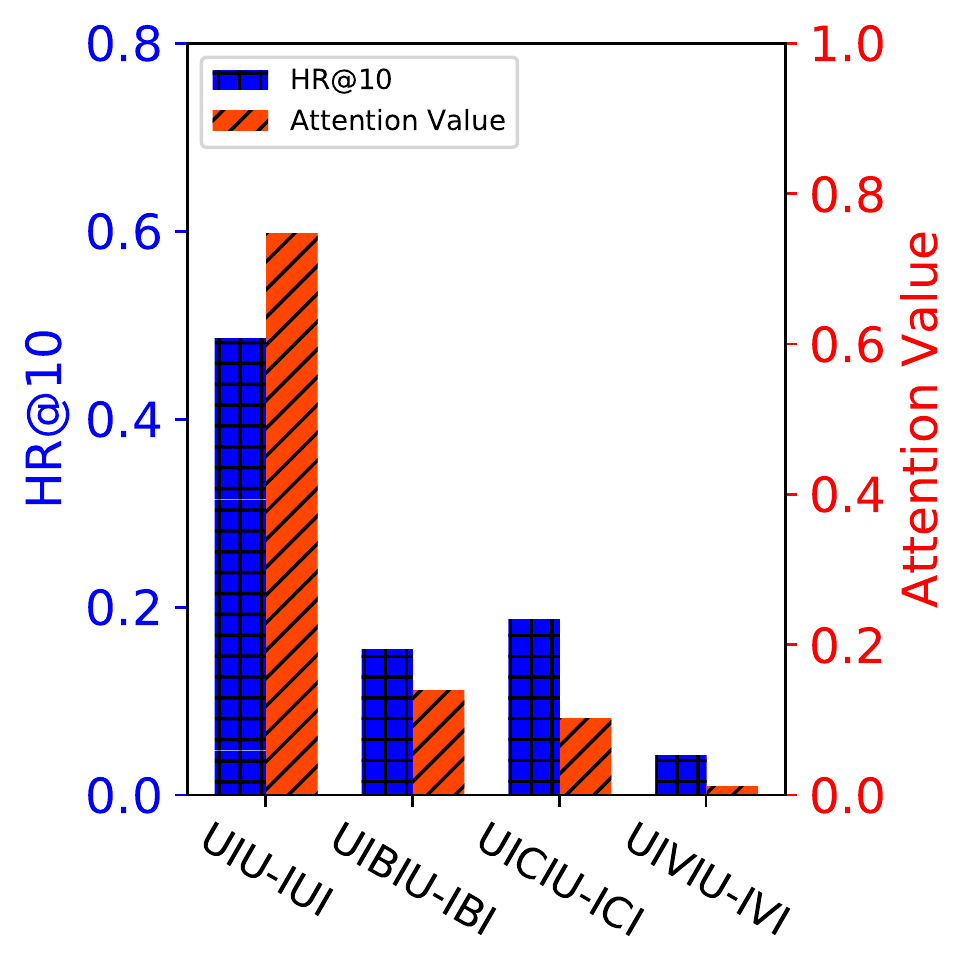}
      \caption{Amazon: NeuACF++}
      \label{fig:amazon_att}
  \end{subfigure}
  \caption{Attention value analysis}
  \label{fig:att}
\end{figure}
\subsubsection{Analysis on Attention}

\begin{figure}
  \centering
  \begin{subfigure}[ht]{0.25\textwidth}
      \centering
      \includegraphics[width=1.6in]{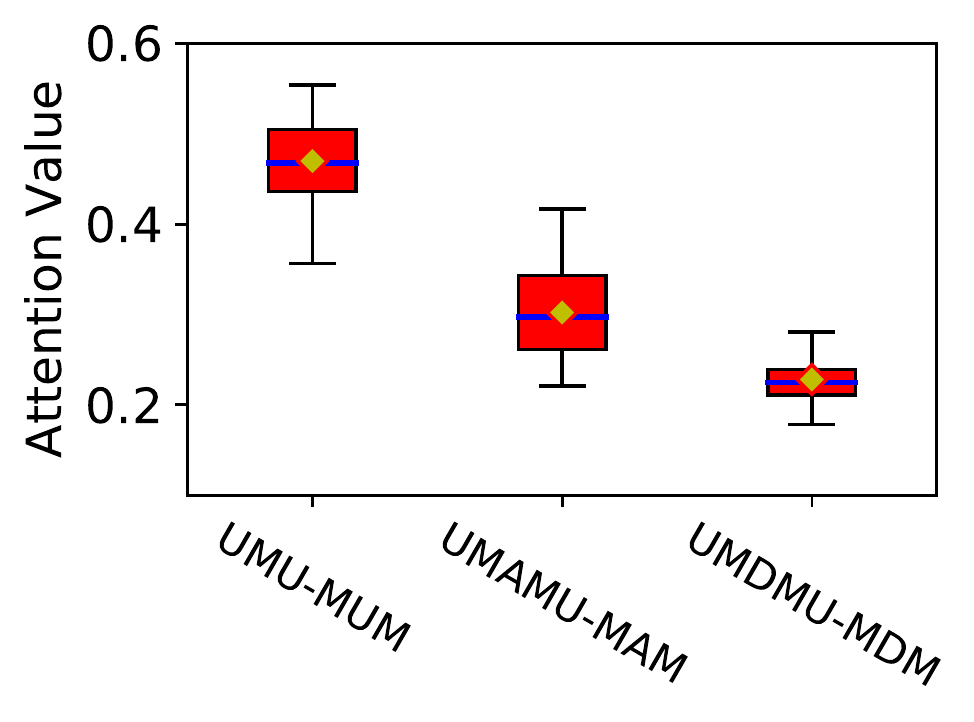}
      \caption{ML100k: NeuACF}
      \label{fig:ml-att_NeuACF_box}
  \end{subfigure}~
  \centering
  \begin{subfigure}[ht]{0.25\textwidth}
      \centering
      \includegraphics[width=1.6in]{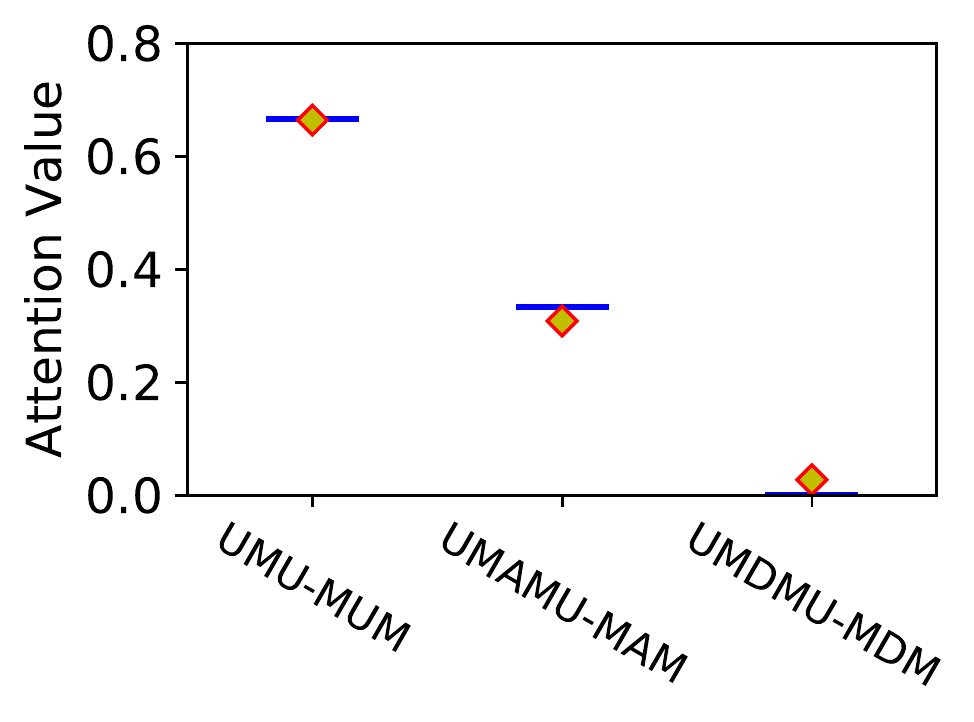}
      \caption{ML100k: NeuACF++}
      \label{fig:ml-att_box}
  \end{subfigure}
  
  \caption{The distribution of attention weights of NeuACF and NeuACF++  on the datasets.}
  \label{fig:att_box}
\end{figure}

In order to investigate that whether the attention values learned from our proposed models NeuACF and NeuACF++ are meaningful, we explore the correlation between the attention values and the recommendation performance of the corresponding meta-path. Generally, we aim to check whether the recommendation performance with one meta-path will is better when the attention value of this meta-path is larger. 

To this end, we conduct experiments to analyze the distribution with attention values and the recommendation performance of single meta-path. 
Specifically, we can obtain the attention value in each aspect for a user based on NeuACF and NeuACF++, and then we are able to average all the attention values for all the users to obtain the final attention value of the aspect. Also, we can get the recommendation results only based on this aspect. So for one aspect, we are able to check the correlation between its recommendation performance and its attention value. Bascially, the better results usually imply that this aspect is more important to the recommendation task, and therefore, this aspect should have larger attention value. We perform experiments with NeuACF and NeuACF++ models respectively. For example, in ML100k dataset, we can obtain three attention values from three different aspect latent factors $UMU$-$MUM$, $UMAMU$-$MAM$, and $UMDMU$-$MDM$ by NeuACF++. 
We present the result of ``Attention Value" and the corresponding single meta-path recommendation results ``HR@10" in Figure~\ref{fig:att}.

One can observe that the attention values of different aspects vary significantly. If the recommendation performance of one meta-path is higher, the corresponding attention value trends to be larger. 
Intuitively, this indicates that the aspect information plays a vital role in recommendation, and ``Average'' is insufficient to fuse different aspect-level latent factors. 
Another interesting observation is that though the distributions of attention values in different datasets are extremely different, the purchase history (e.g. $UMU$-$MUM$ and $UIU$-$IUI$) always takes a large proportion. 
This is consistent with the results in Section~\ref{sec:single_aspect}, suggesting that purchase history usually contains the most valuable information. 

We also present the distribution of attention weights of NeuACF and NeuACF++ on the Movielens dataset in Figure~\ref{fig:att_box}. Figure~\ref{fig:att_box} indicates that the attention values of different aspects are very different and we can find that attention values of NeuACF++ which adopts self-attention are more stable than NeuACF. The reason of this observation is that the self-attention mechanism is more powerful than vanilla attention network to capture the aspect information and  assign more reasonable attention weights to different aspects.

\subsubsection{Visualization of Different Aspect-level Latent Factors}
\begin{figure}[htb]
  \centering
  \begin{subfigure}[]{0.5\textwidth}
      \centering
      \includegraphics[height=0.8in, width=3.0in]{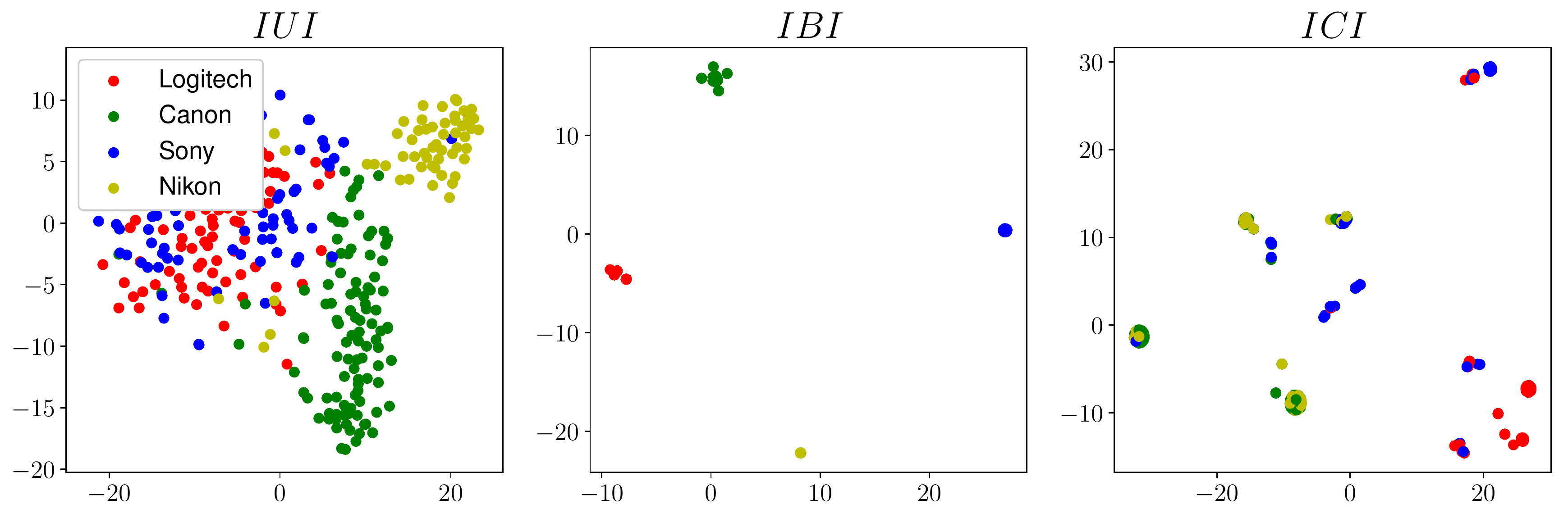}
      \caption{t-SNE embedding with brand labels }
      \label{fig:tsne-b}
  \end{subfigure}%
  
  \centering
  \begin{subfigure}[]{0.5\textwidth}
      \centering
      \includegraphics[height=0.9in, width=3.0in]{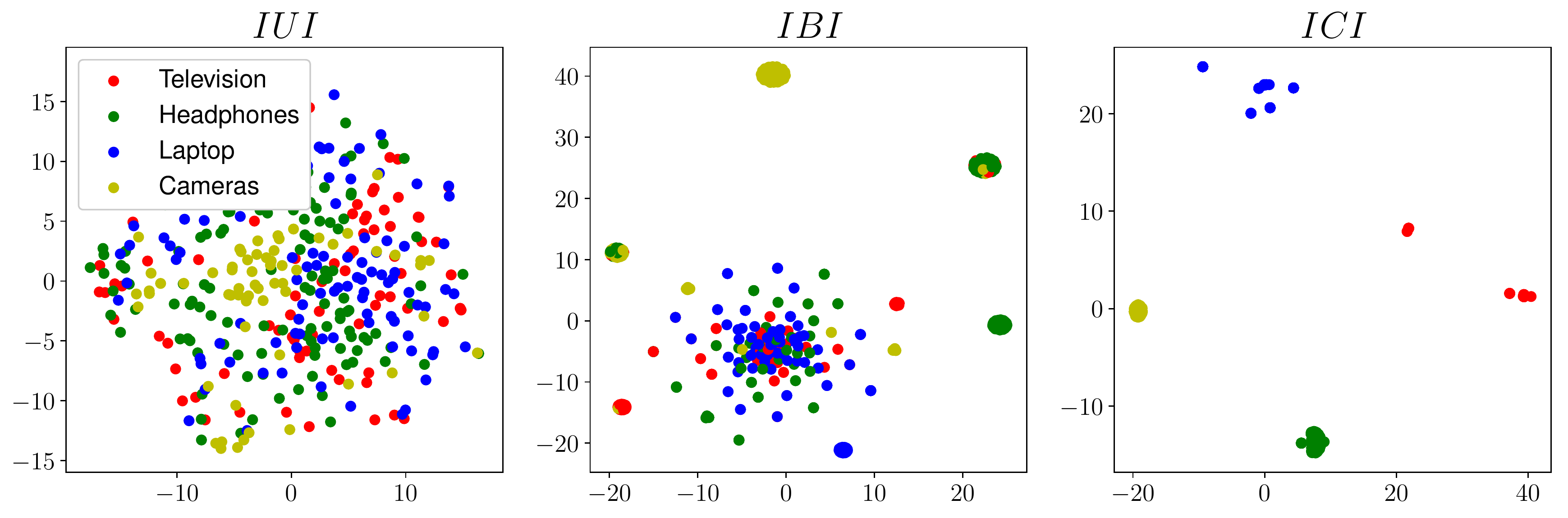}
      \caption{t-SNE embedding with Category labels }
      \label{fig:tsne-c}
  \end{subfigure}%
  \caption{t-SNE embedding with different labels of the learned latent factors of items for Amazon.}
  \label{fig:tsne}
\end{figure}

In our model, we aim to learn the aspect-level latent factors from different meta-paths. For example, we expect that the brand-aspect latent factor $\bm{v}_{j}^{B}$ for item $I_{j}$ can be learned from the meta-path $IBI$, and the category-aspect latent factor $\bm{v}_j^{C}$ from the meta-path $ICI$. To intuitively show whether NeuACF performs well on this task, we visualize the learned aspect-level latent factors on the Amazon dataset. We apply t-SNE~\cite{maaten2008visualizing} to embed the high-dimensional aspect-level latent factors into a 2-dimensional space, and then visualize each item as a point in a two-dimensional space.

Figure~\ref{fig:tsne-b} shows the embedding results for four famous electronics Brand: \emph{Logitech}, \emph{Canon}, \emph{Sony}, and \emph{Nikon}. One can observe that the brand-aspect latent factors can clearly separate the four brands, while the history-aspect and category-aspect latent factors are mixed with each other. It demonstrates the meta-path $IBI$ can learn a good brand-aspect latent factors. Similarly, in Figure~\ref{fig:tsne-c}, only the category-aspect latent factors learned from the meta-path $ICI$ clearly separate the items of different categories including \emph{Television}, \emph{Headphones}, \emph{Laptop} and \emph{Cameras}. The results demonstrate that the aspect-level latent factors of items learned by NeuACF can indeed capture the aspect characteristics of items.

\subsubsection{Parameter Study}
\begin{figure}
  \centering
  \begin{subfigure}[ht]{0.25\textwidth}
      \centering
      \includegraphics[height=1.1in]{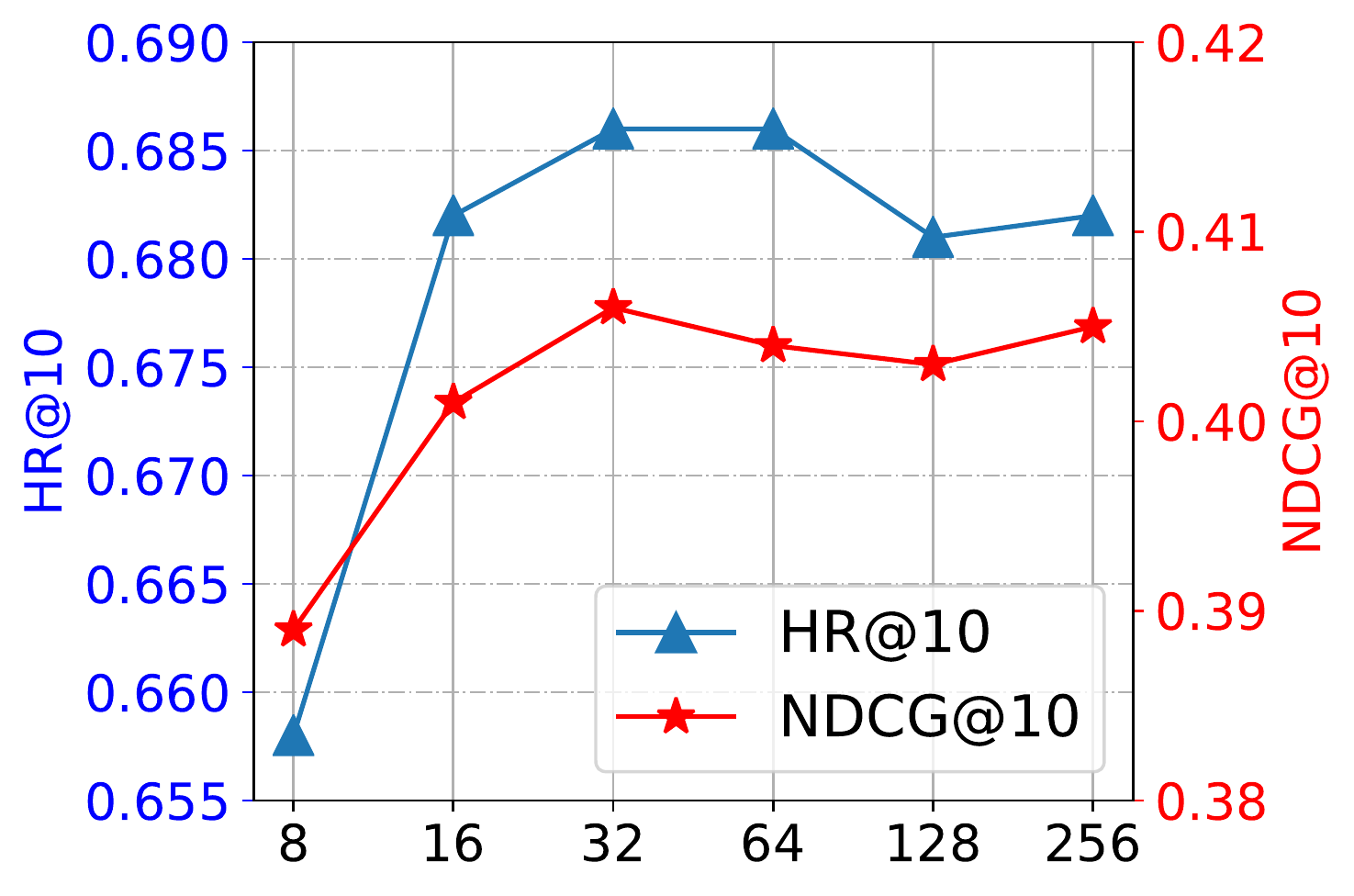}
      \caption{ML100k: NeuACF}
      \label{fig:ml-dimension_NeuACF}
  \end{subfigure}~
  \centering
  \begin{subfigure}[ht]{0.25\textwidth}
      \centering
      \includegraphics[height=1.1in]{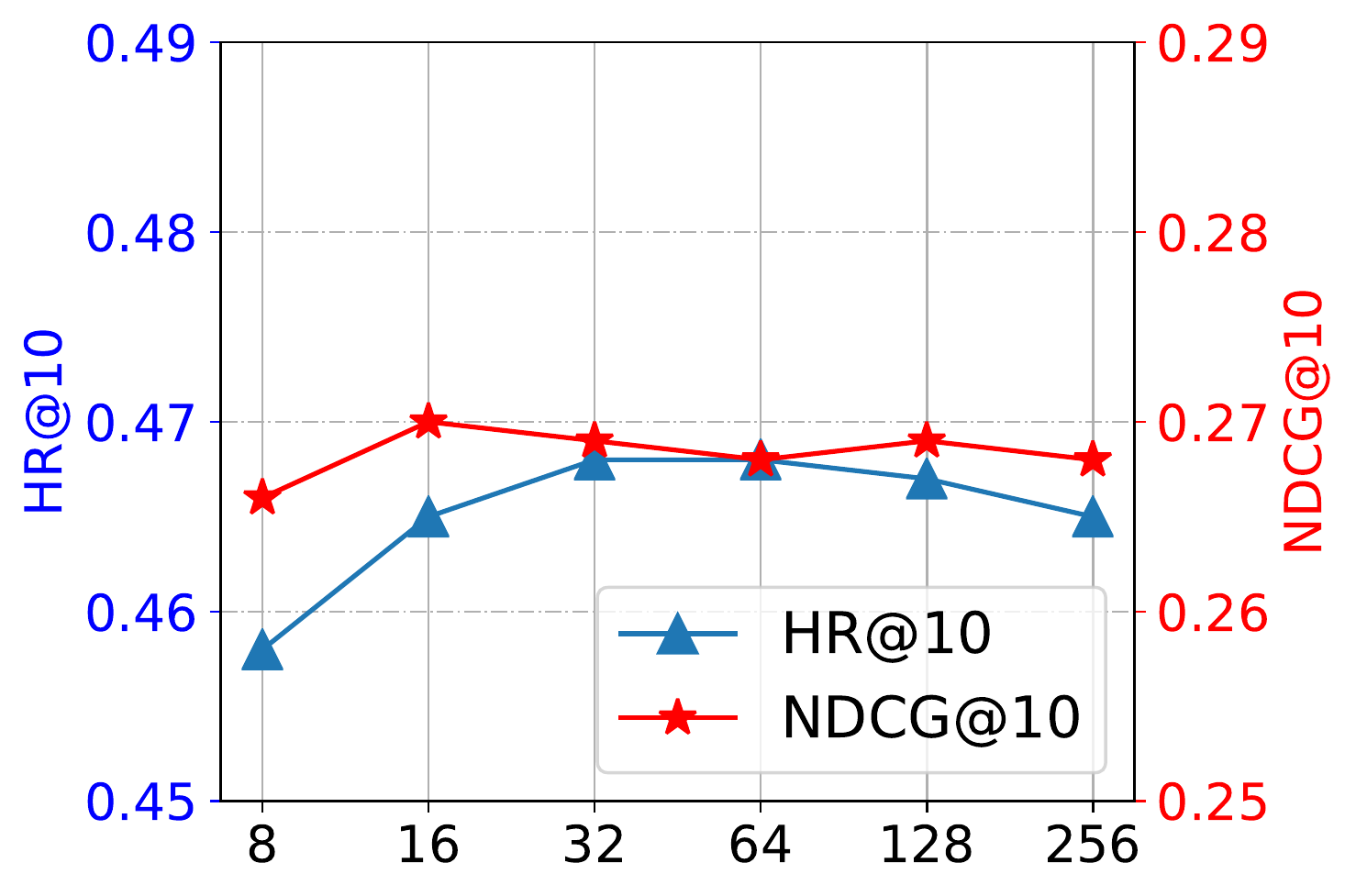}
      \caption{Amazon: NeuACF}
      \label{fig:amazon-dimension_NeuACF}
  \end{subfigure}
  \centering
  \begin{subfigure}[ht]{0.25\textwidth}
      \centering
      \includegraphics[height=1.1in]{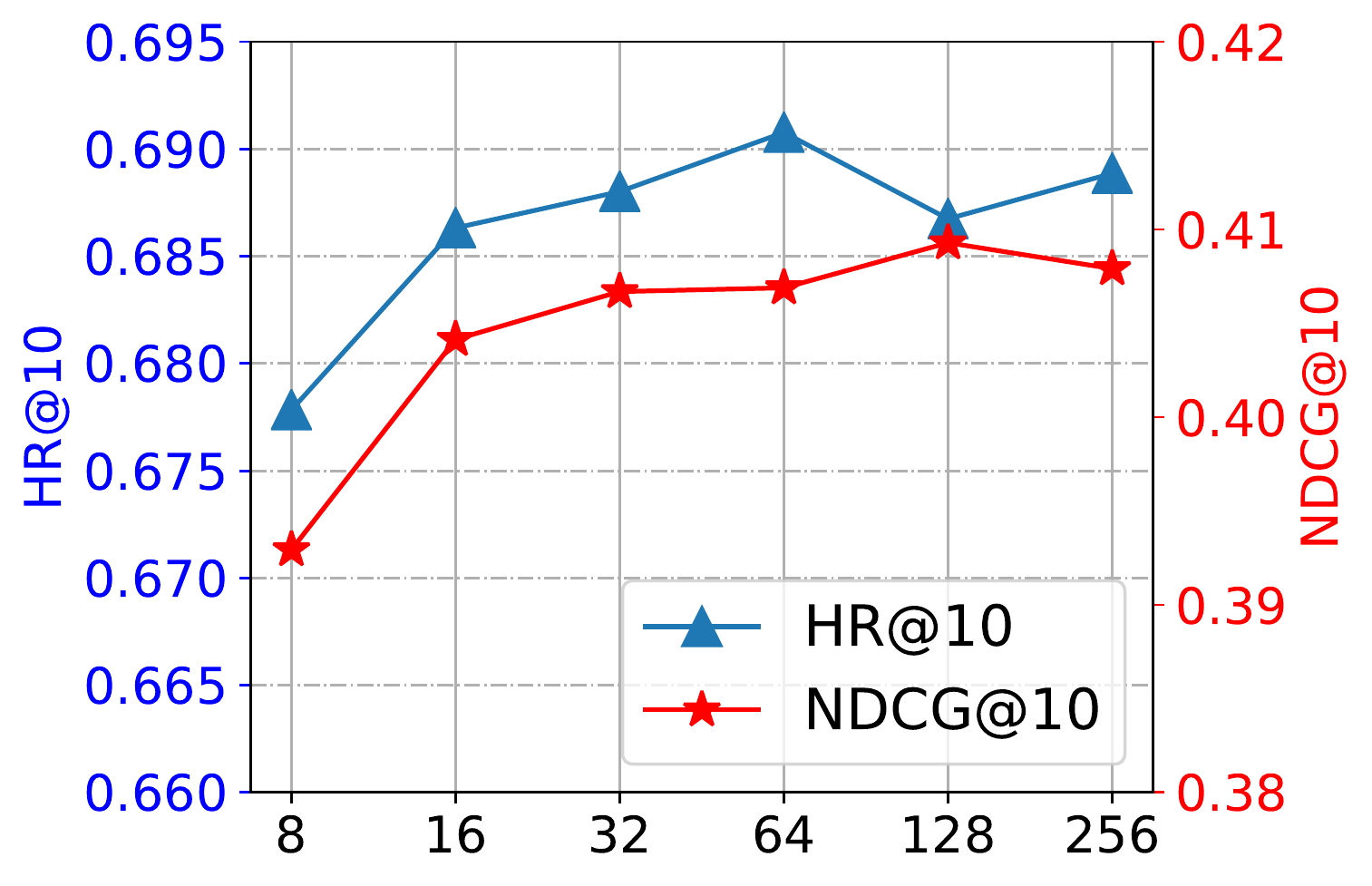}
      \caption{ML100k: NeuACF++}
      \label{fig:ml-dimension}
  \end{subfigure}~
  \centering
  \begin{subfigure}[ht]{0.25\textwidth}
      \centering
      \includegraphics[height=1.1in]{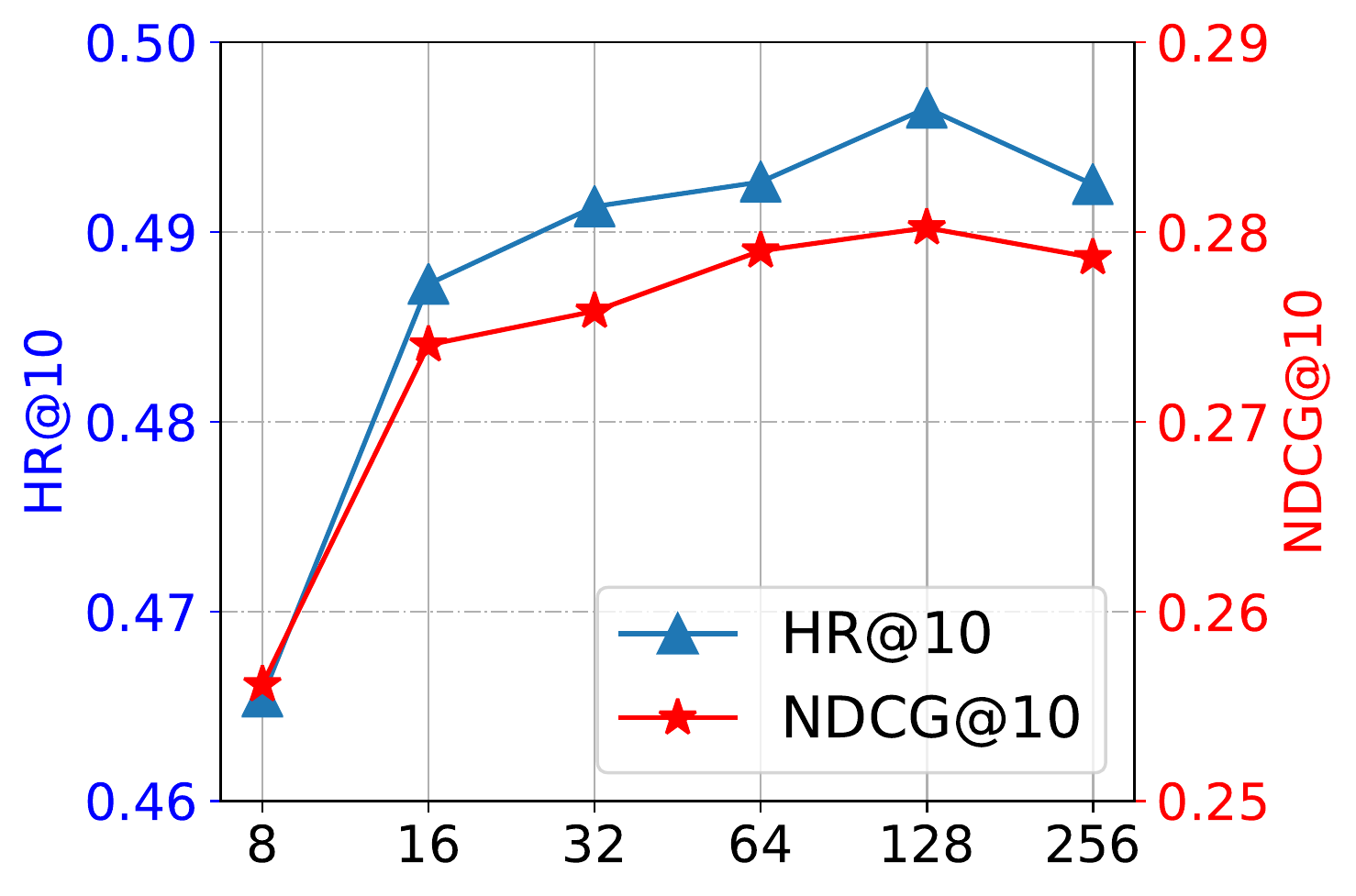}
      \caption{Amazon: NeuACF++}
      \label{fig:amazon-dimension}
  \end{subfigure}
  \caption{Performance with different dimensions of latent factors.}
  \label{fig:dimension}
\end{figure}

\textbf{Effect of the Latent Factor Dimensions}. In the latent factor models, the dimension of the latent factors may have a vital impact on the performance of recommendation. Thus we study the effect of the latent factor dimension learned from the last MLP layer in our proposed model NeuACF and NeuACF++. We conduct the experiment on a three-layer model, and set the dimensions of the latent factors increasing from 8 to 256. The results on the ML100k and Amazon datasets are shown in Figure~\ref{fig:dimension}. 
Figure~\ref{fig:ml-dimension_NeuACF} and Figure~\ref{fig:amazon-dimension_NeuACF} illustrate the performance curve with different numbers of dimensions of NeuACF. 
One can see that on both datasets the performance first increases with the increase of the dimension, and the best performance is achieved at round 16-32. Then the performance drops if the dimension further increases. 
Similarly, Figure~\ref{fig:ml-dimension} and Figure~\ref{fig:amazon-dimension} show the results of NeuACF++.
We can observe that the best performance of NeuACF++ is achieved at round 64 of ML100K and 128 of Amazon. Generally speaking, a small dimension of latent factors is insufficient to capture the complex relationship of users and items. 

\begin{figure}
  \centering
  \begin{subfigure}[ht]{0.25\textwidth}
      \centering
      \includegraphics[height=1.1in]{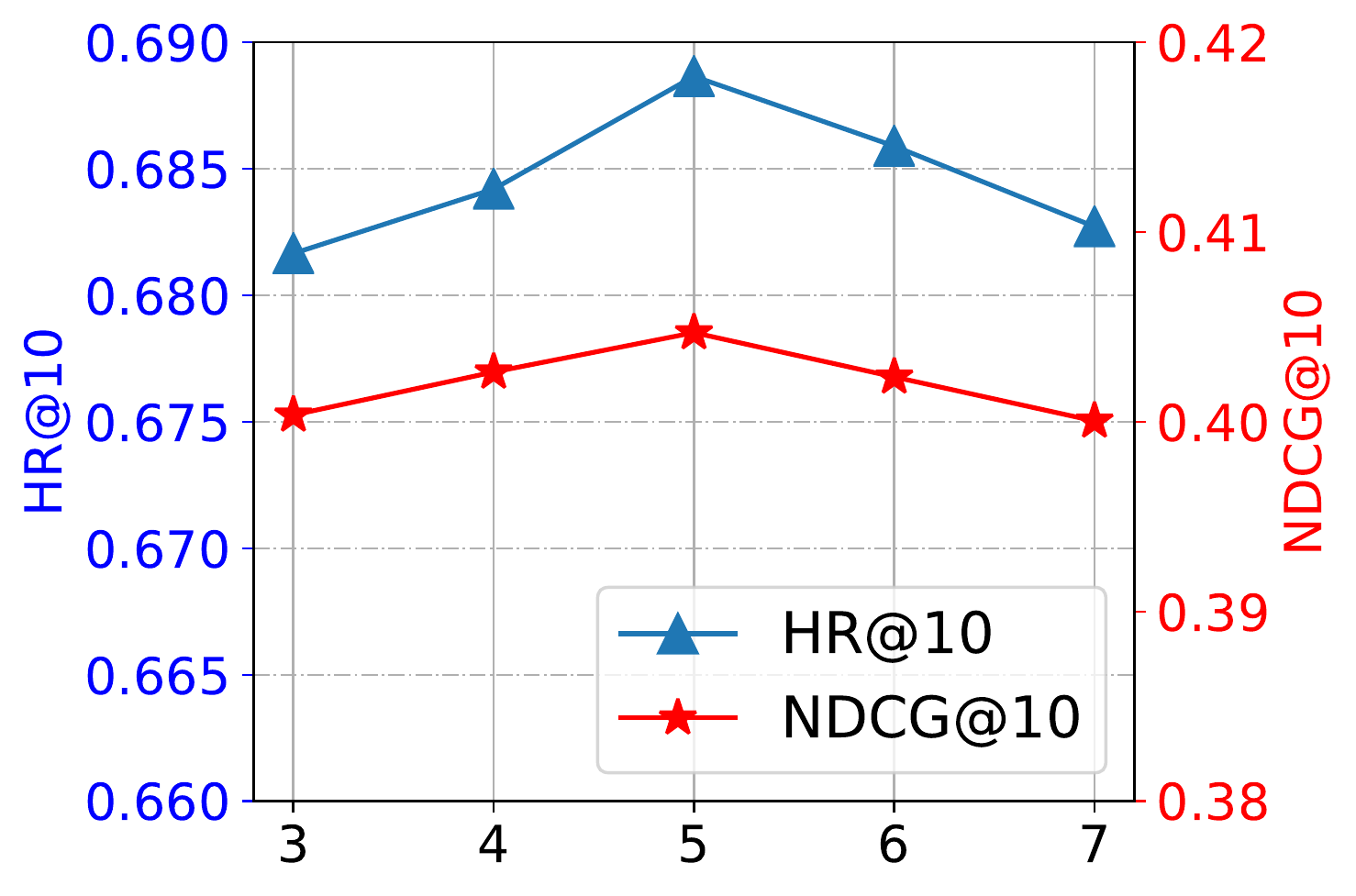}
      \caption{ML100k}
      \label{fig:ml-hidden-num}
  \end{subfigure}~
  \centering
  \begin{subfigure}[ht]{0.25\textwidth}
      \centering
      \includegraphics[height=1.1in]{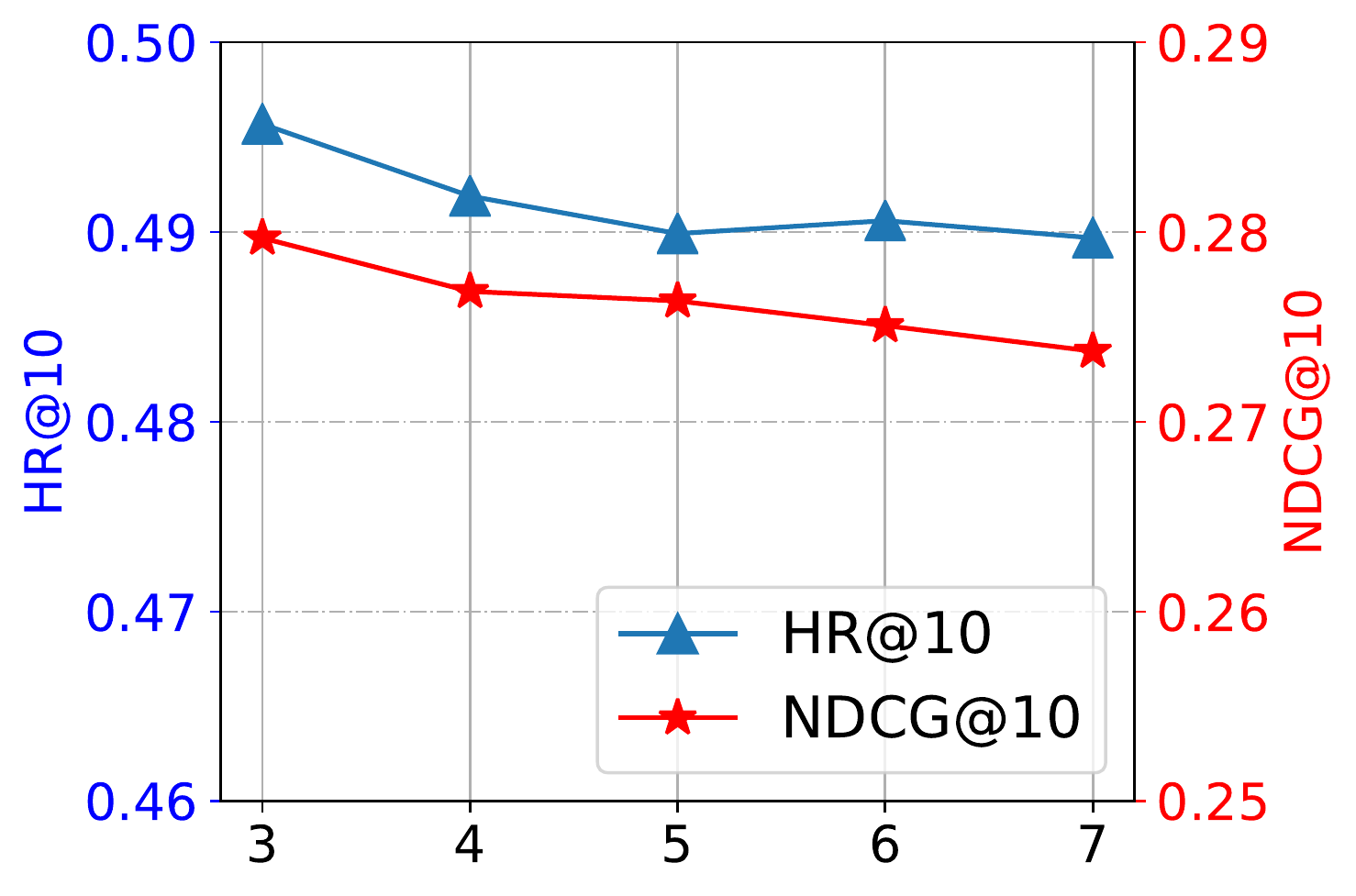}
      \caption{Amazon}
      \label{fig:amazon-hidden-num}
  \end{subfigure}
  \caption{Performance with different numbers of hidden layers.}
  \label{fig:layers}
\end{figure}
\textbf{Effect of Network Hidden Layers}.As the number of hidden layers can usually affect the performance of deep models, we investigate the effect of the number of network hidden layers on our model NeuACF++. We set the number of hidden layers of NeuACF++ from 3 to 7, and the number of hidden neurons of each layer is set up to 64. The results are illustrated in Figure~\ref{fig:layers}. As can be seen from Figure~\ref{fig:ml-hidden-num}, the performance of ML100k dataset first increases with the increase of hidden layers. The best performance is achieved when hidden layers is 5, and then the performance decreases. 
The performance of NeuACF++ decreases slightly when hidden layers increase in Amazon dataset. The best performance is achieved when hidden layers is 3. The reason may be that a three-layer neural network model is capable to characterize the aspect latent factors in Amazon dataset. When the number of hidden layers increase, the model may be over-fitting. From both cases, we can find that the best depth of our model is about 3 layers. Moreover, the slightly degradation may also demonstrate that it is hard for the deep model to learn the identity mapping~\cite{he2016deep}.

\begin{figure}
  \centering
  \begin{subfigure}[ht]{0.25\textwidth}
      \centering
      \includegraphics[height=1.1in]{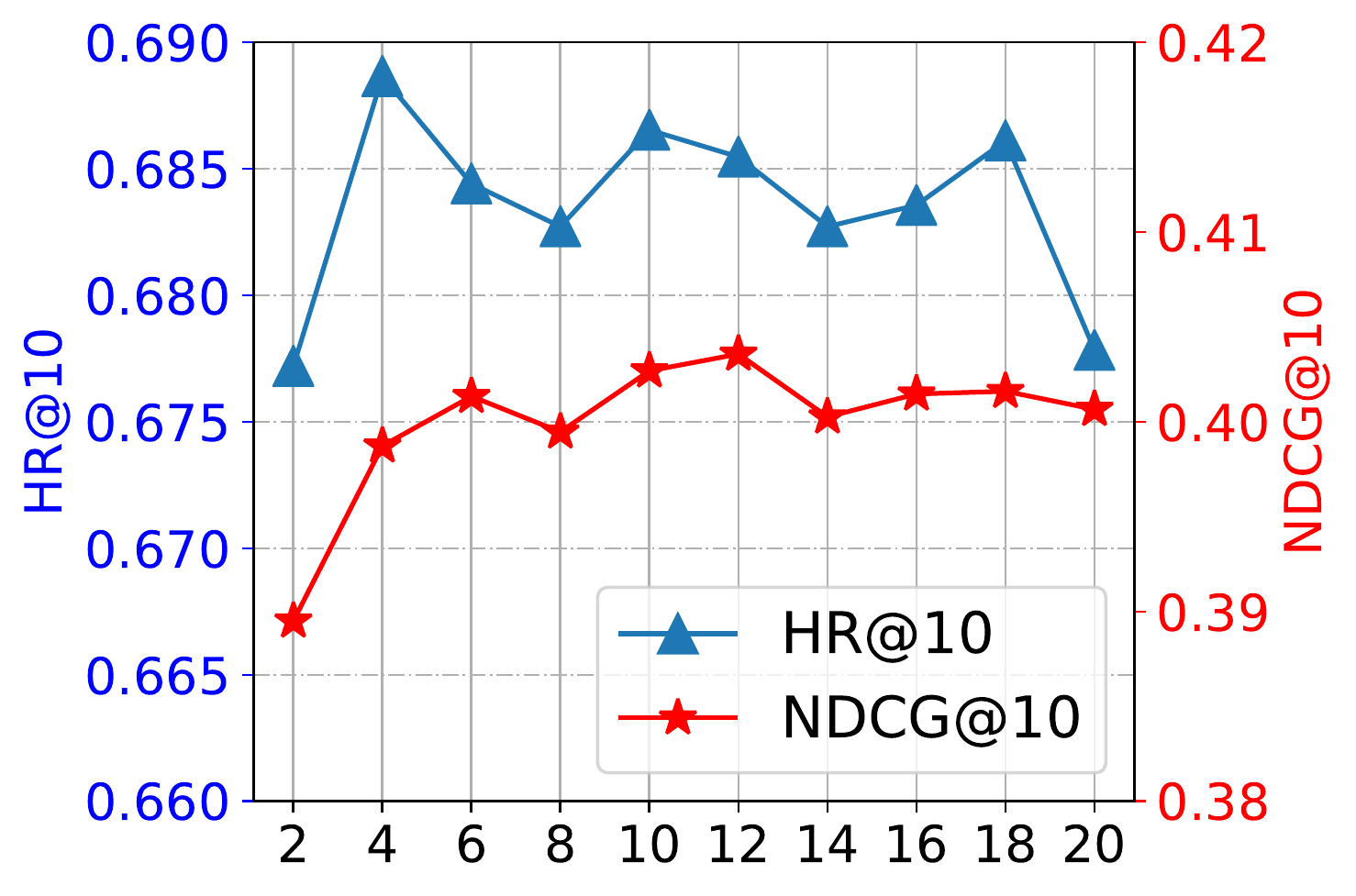}
      \caption{ML100k}
      \label{fig:ml-neg}
  \end{subfigure}~
  \centering
  \begin{subfigure}[ht]{0.25\textwidth}
      \centering
      \includegraphics[height=1.1in]{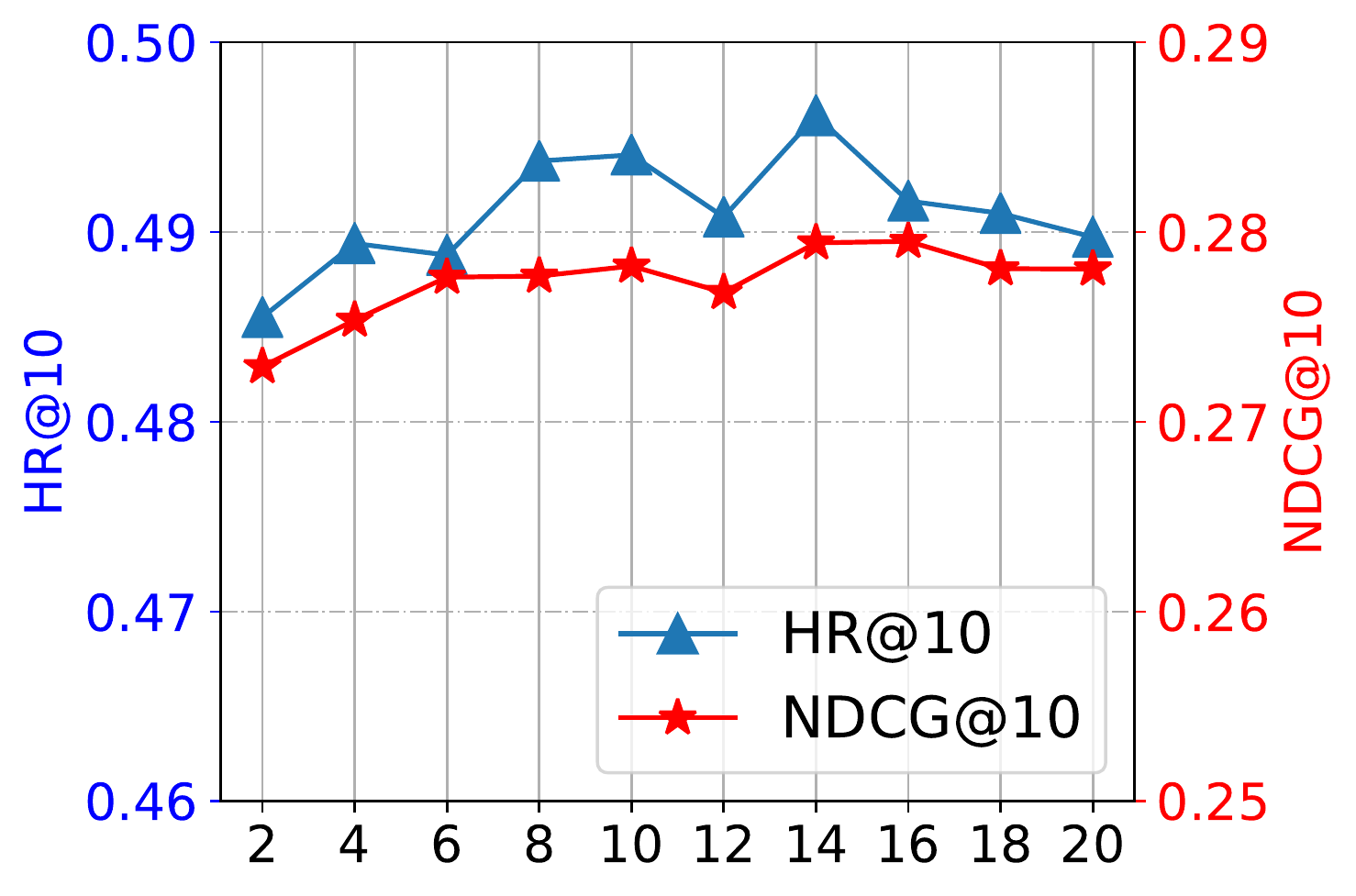}
      \caption{Amazon}
      \label{fig:amazon-neg}
  \end{subfigure}
  \caption{Performance with different number of negative samples.}
  \label{fig:neg}
\end{figure}
\textbf{Effect of Negative Sampling Ratio}. As mentioned above, negative sampling is an effective way to train the neural network model instead of using the whole user-item interactions. 
To illustrate the impact of different negative sampling ratios for NeuACF++ model, we conduct experiments with different negative sampling ratios. The results are shown in Figure~\ref{fig:neg}. The experiments are preformed with the number of negative sampling from 2 to 20 and the increase step is 2. Firstly, Figure~\ref{fig:neg} shows that the number of negative sampling has a significant impact on the model performance. In the ML100k dataset, it demonstrates that less ($\le 4$) negative samples per positive instance is insufficient to achieve optimal performance. It also reveals that setting the sampling ratio too huge ($\ge 10$) may hurt the performance. In Amazon dataset, generally, the performance increases when the number of negative sampling increases. This is probably because of the data sparsity. Table~\ref{tab:statistics} shows that the sparsity of Amazon dataset is about 10 times than ML100k datasets. That means that when the number of negative sampling is 6 in ML100k, there are about 30\% user-item interactions are utilized for the training process. However, even the number of negative sampling is 20 in Amazon dataset, there are only 10\% user-item interactions.

\section{Conclusion}
\label{sec:conclusion}
In this paper, we explore aspect-level information for collaborative filtering. We first propose a novel neural network based aspect-level collaborative filtering model (NeuACF) based on different-aspect features extracted from heterogeneous network with meta-paths. NeuACF is able to learn the aspect-level latent factors and then fuses them with the attention mechanism. Furthermore, in order to better fuse aspect-level information effectively, we propose NeuACF++ which employs the self-attention mechanism to learn the importance of different aspects.
Extensive evaluations demonstrate the superior performance of NeuACF and NeuACF++. 

In this paper, we mainly focus on fusing the latent factors learned in the last layer of the neural network. In the future, we aim to explore new attention mechanism which is able to consider all the latent factor information in all the network layers, so that we can capture more complete information. Moreover, since retraining the model is time-consuming and expensive for new meta-paths, another future work is to design a effective mechanisms to share the neural network which has been learned by before the aspect-level latent factors.

% %% The file named.bst is a bibliography style file for BibTeX 0.99c
% \clearpage
% \bibliographystyle{named}
% \bibliography{paper}

% \end{document}

% \ifCLASSOPTIONcompsoc
% \IEEEraisesectionheading{\section{Introduction}\label{sec:introduction}}
% \else
% \section{Introduction}
% \label{sec:introduction}
% \fi

% or


% \appendices

% % you can choose not to have a title for an appendix
% % if you want by leaving the argument blank
% \section{}

\ifCLASSOPTIONcompsoc
  \section*{Acknowledgments}
This work is supported in part by the National Natural Science Foundation of China (No. 61532006, 61772082, 61702296, 61602237), the National Key Research and Development Program of China (2017YFB0803304), the Beijing Municipal Natural Science Foundation (4182043), and the 2019 CCF-Tencent Open Research Fund. This work is also supported in part by NSF under grants III-1526499, III-1763325, III-1909323, SaTC-1930941, and CNS-1626432.
  
\else
  \section*{Acknowledgment}

\fi

\ifCLASSOPTIONcaptionsoff
  \newpage
\fi



\bibliographystyle{IEEEtran}
\bibliography{IEEEabrv,./paper.bib}
% \bibliography{./ref.bib}


% H.~Kopka and P.~W. Daly, \emph{A Guide to {\LaTeX}}, 3rd~ed.\hskip 1em plus

% \end{thebibliography}


\vspace{-70pt}
\begin{IEEEbiography}
[{\includegraphics[width=0.8in
 ,clip,keepaspectratio
 ]{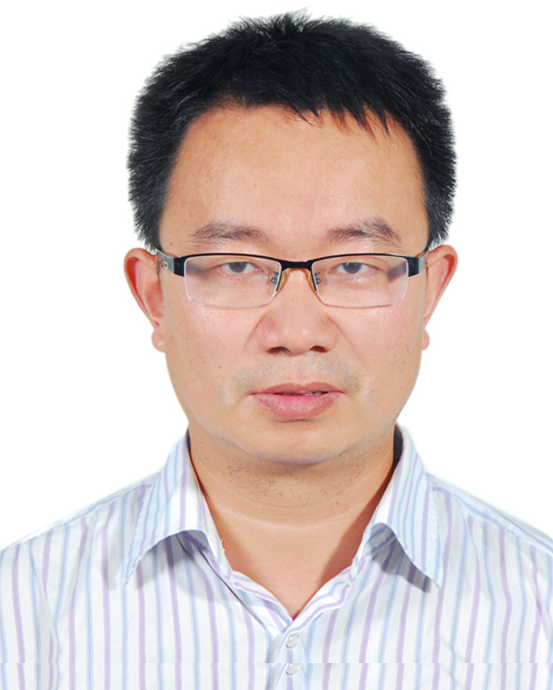}}]
{Chuan Shi}
received the B.S. degree from the Jilin University in 2001, the M.S. degree from the Wuhan University in 2004, and Ph.D. degree from the ICT of Chinese Academic of Sciences in 2007. He joined the Beijing University of Posts and Telecommunications as a lecturer in 2007, and is a professor and deputy director of Beijing Key Lab Intelligent Telecommunications Software and Multimedia at present. His research interests are in data mining, machine learning, and evolutionary computing. He has published more than 60 papers in refereed journals and conferences, such as TKDE, KAIS, TIST, KDD, IJCAI.
\end{IEEEbiography}

\vspace{-55pt}
\begin{IEEEbiography}[{\includegraphics[width=0.8in
 ,clip,keepaspectratio
 ]{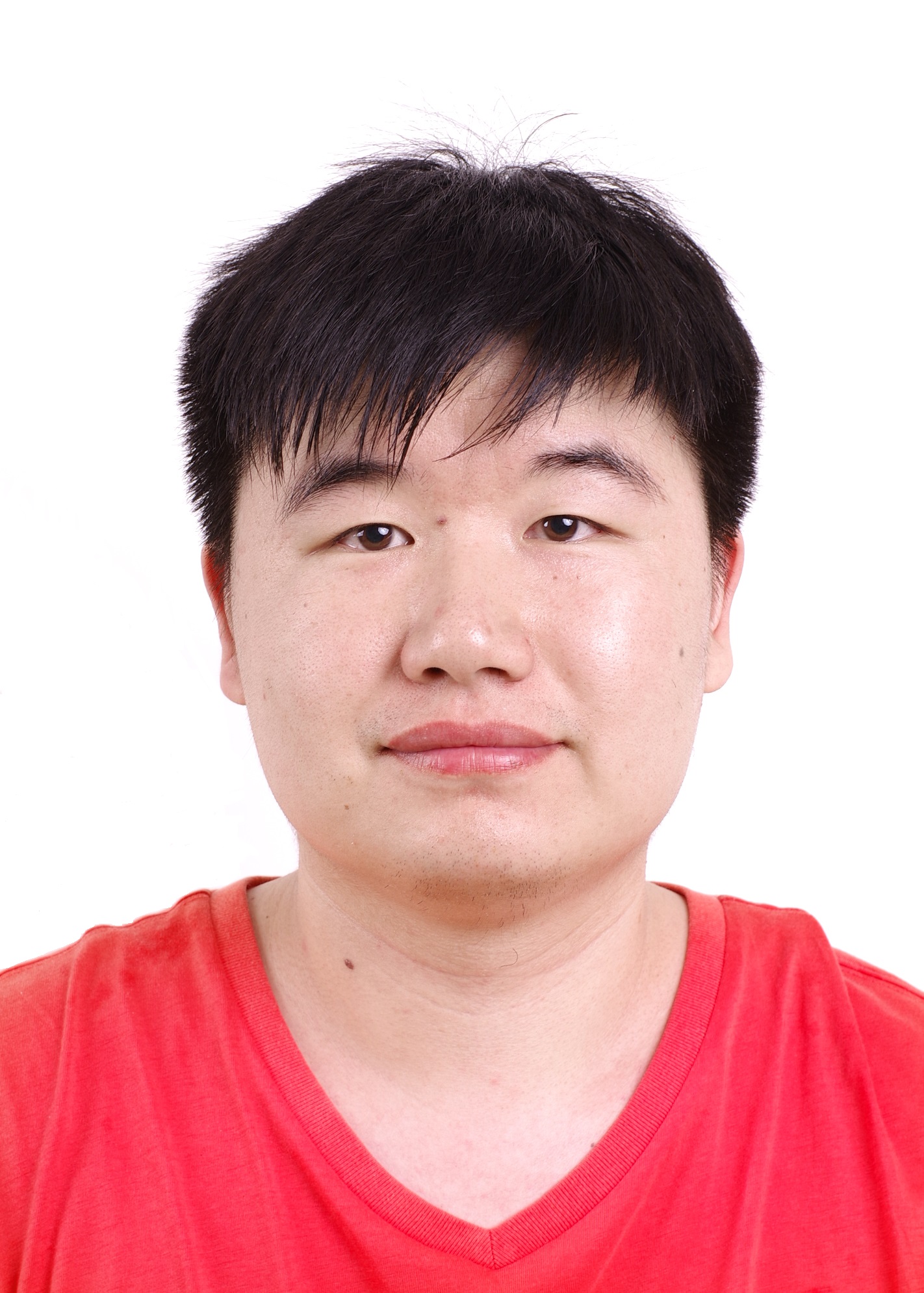}}]{Xiaotian Han}
 is currently a Ph.D. student at Texas A\&M University.
He received the M.S. degree from Beijing University of Posts and Telecommunications in Computer Science, and B.S. degree from Shandong University. His research interests are social network and recommender system.
\end{IEEEbiography}

\vspace{-50pt}
\begin{IEEEbiography}
[{\includegraphics[width=0.8in
 ,clip,keepaspectratio
 ]{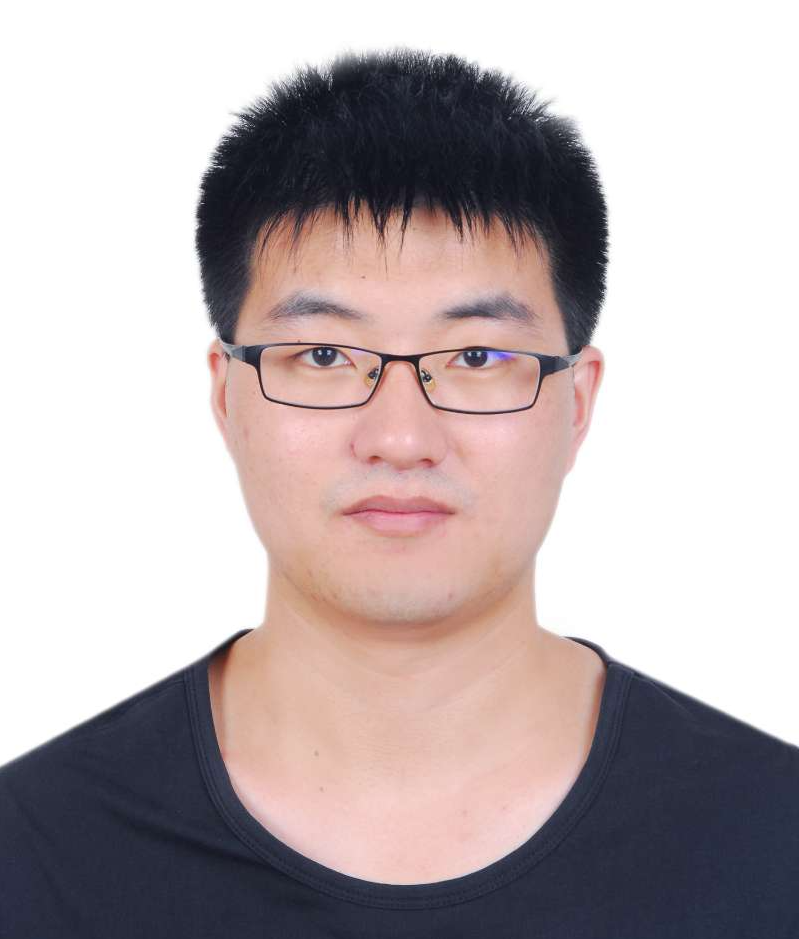}}]
{Li Song}
is currently an algorithm engineer at JD.
He received the M.S. degree from Beijing University of Posts and Telecommunications in 2019, and B.S. degree from North China Electric Power University.  His research interests are recommender system and trajectory data mining.
\end{IEEEbiography}

\vspace{-50pt}
\begin{IEEEbiography}
[{\includegraphics[width=0.8in
 ,clip,keepaspectratio
 ]{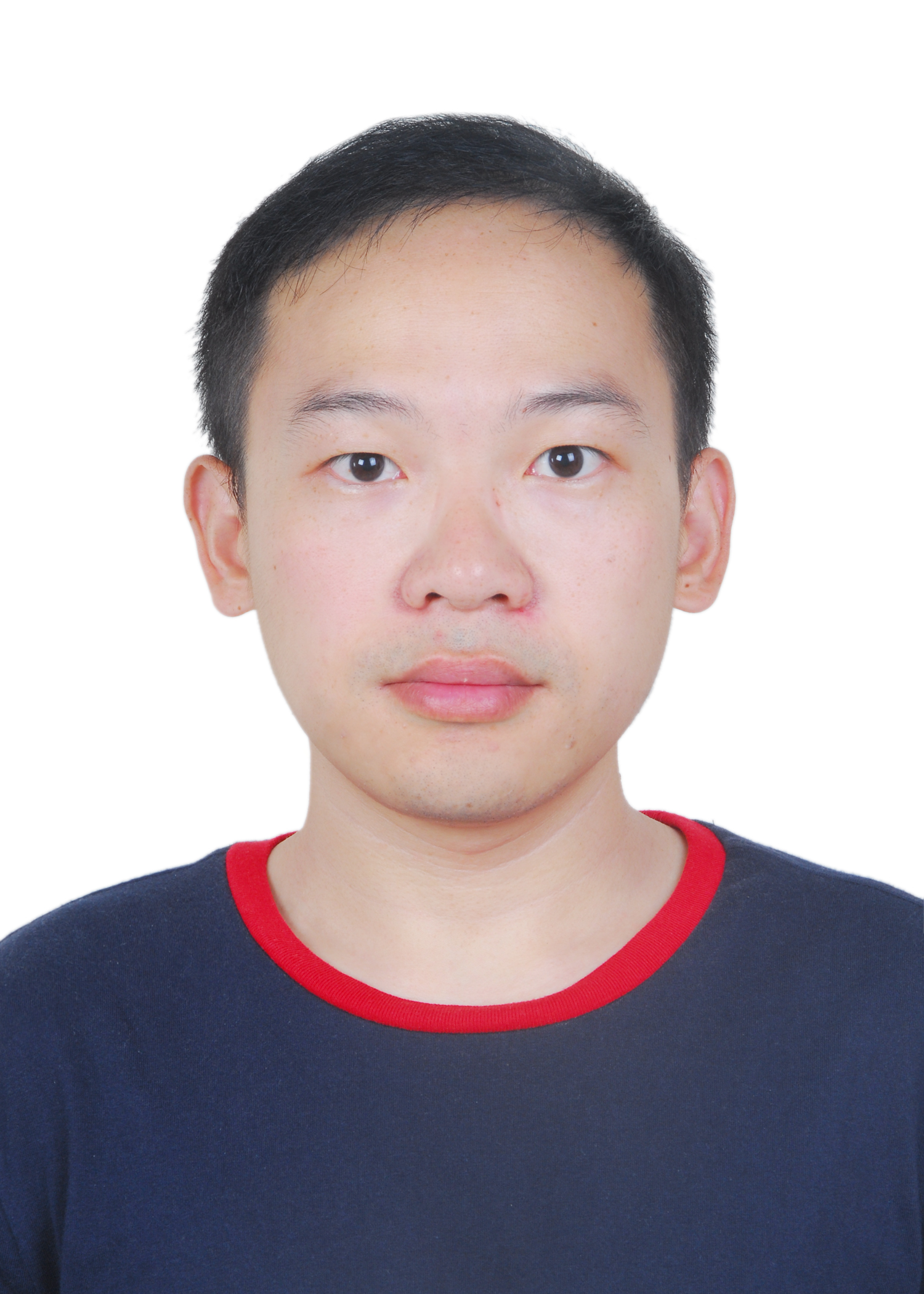}}]
{Xiao Wang}
 is currently an Assistant Professor at Beijing University of Posts and Telecommunications, Beijing, China. Prior to that, he was a postdoctoral researcher in Department of Computer Science and Technology, Tsinghua University, Beijing, China. He received his Ph.D. degree from the School of Computer Science and Technology, Tianjin University, Tianjin, China, in 2016.  He got the China Scholarship Council Fellowship in 2014 and visited Washington University in St. Louis, USA, as a joint training student from Nov. 2014 to Nov. 2015. His current research interests include data mining, social network analysis, and machine learning. Until now, he has published more than 30 papers in conferences such as AAAI, IJCAI, etc. and journals such as IEEE Trans. on Cybernetics, IEEE Trans. on Knowledge Discovery and Engineering, etc. Now his research is sponsored by National Science Foundation of China.
\end{IEEEbiography}

\vspace{-35pt}
\begin{IEEEbiography}[{\includegraphics[width=0.8in
 ,clip,keepaspectratio
 ]{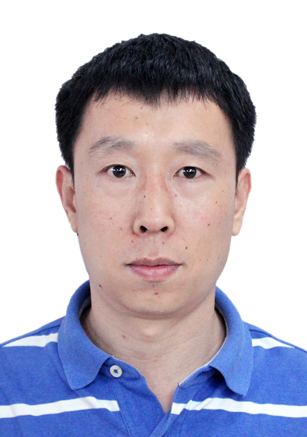}}]
 {Senzhang Wang} received the B.Sc. and Ph.D. degree in Southeast University, Nanjing, China in 2009 and Beihang University, Beijing, China in 2016 respectively. He is currently an associate processor at College of Computer Science and Technology, Nanjing University of Aeronautics and Astronautics, and also a “Hong Kong Scholar” Postdoc Fellow at Department of Computing, The Hong Kong Polytechnic University. His main research focus is on data mining, social computing, and urban computing. He has published more than 70 referred conference and journal papers.
\end{IEEEbiography}

\vspace{-35pt}
\begin{IEEEbiography}[{\includegraphics[width=0.8in
 ,clip,keepaspectratio
 ]{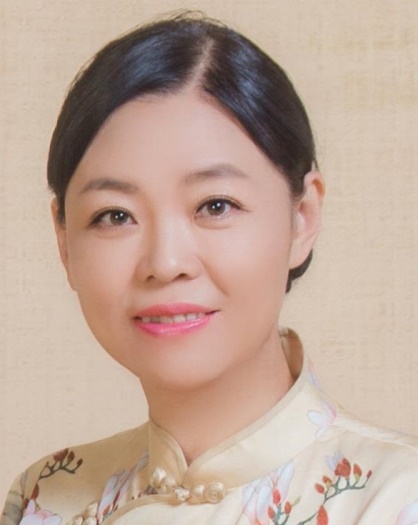}}]
 {Junping Du} received the PhD degree in computer science from the University of Science and Technology Beijing (USTB), and then held a postdoc fellowship in the Department of Computer Science, Tsinghua University, Beijing, China. She joined the School of Computer Science, Beijing University of Posts and Telecommunications (BUPT), in July 2006, where she is currently a professor of computer science. She was a visiting professor in the Department of Computer Science, Aarhus University, Denmark, from September 1996 until September 1997. Her current research interests include artificial intelligence, data mining, intelligent management system development, and computer applications.
\end{IEEEbiography}

\vspace{-35pt}
\begin{IEEEbiography}
[{\includegraphics[width=0.8in
 ,clip,keepaspectratio
 ]{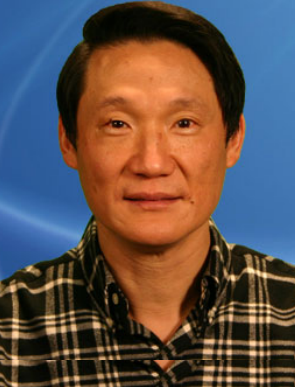}}]
{Philip S. Yu}
received the BS degree in electrical engineering from National Taiwan University, the MS and PhD degrees in electrical engineering from Stanford University, and the MBA degree from New York University. He is a distinguished professor in computer science with the University of Illinois at Chicago and also holds the Wexler chair in information technology. His research interests include big data, including data mining, data stream, database, and privacy. He has published more than 1,000 papers in refereed journals and conferences. He holds or has applied for more than 300 US patents. He is the recipient of the ACM SIGKDD 2016 Innovation Award for his influential research and scientific contributions on mining, fusion, and anonymization of big data, the IEEE Computer Societys 2013 Technical Achievement Award for pioneering and fundamentally innovative contributions to the scalable indexing, querying, searching, mining, and anonymization of big data, and the Research Contributions Award from IEEE ICDM in 2003 for his pioneering contributions to the field of data mining. He also received the ICDM 2013 10-year Highest-Impact Paper Award, and the EDBT Test of Time Award (2014). He was the editor-inchiefs of the ACM Transactions on Knowledge Discovery from Data (2011-2017) and the IEEE Transactions on Knowledge and Data Engineering (2001-2004). He is a fellow of the IEEE and ACM.
\end{IEEEbiography}

%\newpage



%\vfill


\end{document}